 \documentclass[preprint,review,12pt,numbers]{elsarticle}

\usepackage{amsmath,amsfonts,amssymb}
\usepackage{booktabs}  % For better-looking tables
\usepackage{multirow}  % For multi-row cells
\usepackage{epsfig}
\usepackage{float}
\usepackage[margin=0.75in]{geometry}
\usepackage{color,soul}
\usepackage{lineno}

\usepackage[utf8x]{inputenc}
\usepackage[colorinlistoftodos]{todonotes}
\usepackage{csquotes}
\MakeOuterQuote{"}

\usepackage{caption}
\usepackage{amsmath}
\usepackage[ruled,vlined,linesnumbered]{algorithm2e}
\DontPrintSemicolon

\usepackage{subcaption}   % provides subfigure + caption control
\usepackage{booktabs}
\usepackage{siunitx}
\newlength{\steplabelwidth}
\usepackage{tabularx}
\usepackage{booktabs}

\usepackage{hyperref}

\usepackage{makecell}

\begin{document}

\begin{frontmatter}



\title{A causal graph-informed temporal convolution architecture for interpretable retail electricity price forecasting}


\affiliation[inst1]{organization={Risk Assessment and Management of Structural and Infrastructural Systems (RAMSIS) Laboratory, Department of Civil, Environmental, and Geodetic Engineering, The Ohio State University},%Department and Organization 
            city={Columbus},
            postcode={43210}, 
            state={OH},
            country={USA}}

\affiliation[inst2]{organization={John Glenn College of Public Affairs, The Ohio State University},%Department and Organization
            city={Columbus},
            postcode={43210}, 
            state={OH},
            country={USA}}

\author[inst1]{Yufan Ji}
\author[inst1]{Abdollah Shafieezadeh}
\author[inst2]{Noah Dormady}

\begin{abstract}

Retail electricity markets in deregulated systems face significant price volatility and complex interactions with forward and futures products, posing challenges for effective operational decision-making. This study introduces a Causal Graph-Informed Temporal Convolutional Network (CG-TCN), a forecasting architecture that integrates a learned causal graph into a temporal convolutional network via a graph-neural embedding to enhance both forecasting accuracy and interpretability of retail electricity price dynamics. The framework first applies a multi-resolution decomposition to isolate semiannual, quarterly, and monthly trends from high-frequency fluctuations. A causal graph is then discovered over these components and key covariates—including wholesale forward prices and retail contract attributes such as early termination fees—using NOTEARS, with domain constraints that preserve causal directionality and exogeneity. The learned causal structure is encoded as an adjacency embedding that conditions the TCN's convolutions and attention, aligning representation learning with causal pathways. To better capture rare but impactful retail price surges, we use a spike-aware training algorithm that oversamples spike windows and up-weights their errors. Using ten years of daily 12-month fixed-price residential contracts from Ohio’s deregulated market, we find that wholesale forward prices primarily determine long-term retail price trends, whereas contract attributes influence short-term fluctuations. CG-TCN consistently outperforms benchmark models, achieving mean absolute percentage errors of 3.08\%, 3.82\%, and 5.43\% for one-, ten-, and fifteen-step-ahead forecasts of daily retail electricity median prices, respectively. By combining predictive performance with interpretability, CG-TCN provides transparent, policy-relevant insight to support market analytics, consumer protection, regulatory oversight, risk assessment and procurement planning in competitive electricity markets.

\end{abstract}


\end{frontmatter}

%\linenumbers

\section{Introduction}

%% the importance of forecasting retail electricity prices
Retail electricity prices play a critical role in shaping both consumer expenditures and supplier strategy in deregulated markets. For households and small businesses, these prices directly determine energy expenditures and influence adoption of efficiency measures, distributed generation, and demand-response programs \cite{faruqui2013dynamic, dormady2025retail}. For suppliers and policymakers, retail prices serve as signals for market competitiveness, hedging strategies, and the integration of renewable resources \cite{dormady2025efficiency}. Yet retail prices arise from shifting structural drivers—including wholesale forward prices, regulatory interventions, and evolving contract attributes—that generate pronounced volatility and occasional price spikes. Reliable retail price forecasts are therefore essential for regulators tasked with consumer protection and market participants requiring effective procurement planning and risk management. Equally important, models that uncover causal mechanisms behind price formation can inform robust market design, guide investment, and support a transparent energy transition.

Retail electricity prices in deregulated markets are jointly shaped by wholesale cost conditions and retail market design, yet existing studies largely examine these drivers in isolation rather than within a unified analytical framework. In deregulated retail electricity markets, consumers can choose among contracts offered by multiple competitive suppliers, each featuring different terms and conditions. In Ohio, for example, more than twenty suppliers collectively post nearly one hundred residential electricity contracts per day within each utility service territory, across six territories statewide \cite{dormady2025efficiency}. Recent studies show that contract attributes such as early termination fees and monthly fixed charges %significantly influence the offered retail price: 
are systematically related to offered prices—higher fees are often associated with lower per-kWh rates \cite{ji2025renewable}. These contract-level features influence retail prices alongside other more structural wholesale market attributes in the broader market (e.g., NYMEX futures, implied volatility) \cite{dormady2025retail}. Wholesale forward prices shape the costs of suppliers (or marketers) that supply competitive retail service and/or procurements through default service known as the “standard service offer” (SSO). The SSO is the default supply rate charged by suppliers and collected on distribution utility bills for non-shopping customers. The SSO  often serves as a baseline reference for households evaluating contract options, and in most jurisdictions it is even referred to as a comparison price printed on customer bills. \enlargethispage{1\baselineskip}\footnote{This comparison price is commonly referred to as the “Price to Compare” (PTC) in Ohio, New York, Pennsylvania, and several other deregulated retail electricity markets.} Incorporating both contract-level and market-level drivers is therefore essential for understanding how retail electricity prices are formed. This is of critical importance in energy markets because there are very few studies of price formation in retail markets, and no present studies evaluating the combined (wholesale and retail) market attributes influencing retail price formation.

%% forecasting methods 
Although limited to wholesale markets, a growing body of research has advanced electricity price forecasting and explored linkages between market fundamentals and prices \cite{pesenti2025explaining, aggarwal2009electricity, vega2021use}. For example, Wang et al. developed a convolutional neural network–long short-term memory ensembles to forecast day-ahead prices in Japan \cite{wang2024novel}. Cerasa et al. proposed a filtering strategy to enhance forecasting accuracy in U.S. and European markets by systematically identifying and replacing extreme spikes \cite{cerasa2025enhancing}. Kohút et al. employed recurrent neural networks to forecast wholesale prices across European grids \cite{kohut2025unified}. While these approaches achieve acceptable predictive accuracy, they remain largely “black-box” models, offering limited insight into the mechanisms driving price dynamics. This limits robustness under policy or regime changes and weakens their decision value for procurement and regulation.

%% causality in forecasting
As demand grows for interpretability and explainability in machine learning forecasts, researchers have begun to report additional details about model training and feature relevance \cite{pesenti2025explaining, heistrene2024improved}. Yet much of the literature remains correlation-based, optimizing prediction error or reporting feature importance without establishing directional, invariant relationships among drivers. This limits their ability to generalize under regime shifts or policy shocks. Recent evidence shows that embedding causal structures into forecasting models can improve robustness and interpretability \cite{ganesan2019use, miraki2024electricity}. Unlike feature importance, causal analysis can address confounding and distinguish mediators from true causes. To fill this gap, we propose a Causal-Graph-informed Temporal Convolutional Network (CG-TCN) for the more complicated and policy-relevant and consumer-relevant application of retail electricity price forecasting. Our approach first learns a directed causal graph, and then injects that graph into forecasting via a graph-neural embedding that conditions the TCN, %encodes this graph as an inductive bias in the temporal convolutional network, 
aligning representation learning with hypothesized causal pathways.

%% The introduction to NOTEARS.....
A prominent method for causal graph discovery is NOTEARS (Non-combinatorial Optimization via Trace Exponential and Augmented lagRangian for Structure learning) \cite{zheng2018dags}. Traditional approaches such as the Peter–Clark (PC) algorithm (constraint-based) \cite{kalisch2007estimating} or the Greedy Equivalence Search (GES) algorithm (score-based) \cite{chickering2002optimal} often rely on combinatorial searches over graph structures, which scale poorly and can be sensitive to conditional independence tests or scoring heuristics. In contrast, NOTEARS learns a directed acyclic graph (DAG) by solving a smooth, continuous optimization with a differentiable acyclicity constraint, enabling efficient gradient-based optimization while guaranteeing a valid DAG. This property makes NOTEARS particularly attractive for high-dimensional forecasting problems, where correlated drivers and latent common shocks can obscure true directional relationships.

In this study, we employ NOTEARS to infer the causal graph among retail electricity price drivers while imposing economically motivated domain constraints. Specifically, the retail price target is restricted from acting as a parent node, preventing implausible feedback loops in which future prices influence upstream drivers. This restriction ensures that inferred relationships reflect economically meaningful drivers of price formation rather than reverse causation. In addition, exogenous variables—such as wholesale forward prices and time features—are constrained to remain exogenous, disallowing incoming edges from retail or contract-level variables. These constraints encode basic economic structure, reflecting the fact that wholesale market conditions and institutional calendars shape retail pricing decisions but are not themselves determined by individual retail contracts or observed price realizations. By incorporating such domain knowledge, the learned causal graph becomes more stable, interpretable, and economically coherent, filtering out spurious correlations that arise from shared trends or temporal persistence. The resulting structure then serves as an inductive bias for our CG-TCN, aligning temporal representation learning with hypothesized causal mechanisms underlying retail electricity price formation.

%% the function of decomposition
In parallel, time-series decomposition techniques have proven valuable for improving electricity price forecasts, though many applications treat decomposed components as purely statistical artifacts and do not explicitly link them to underlying economic mechanisms. In this study, we employ causal, one-sided multi-resolution decompositions based on Seasonal–Trend decomposition using Loess (STL) and a parametric filter-bank approach, which we refer to as a Temporal Smoothing Network (TSN)—a deterministic, multi-scale bank of causal smoothing filters rather than a learned neural network-to extract horizon-specific components from retail electricity prices in real time. Electricity prices are known to exhibit multi-scale seasonality and structural patterns that correspond to distinct economic drivers—for example, long-term seasonal and trend components have been shown to improve forecasting accuracy when modeled separately from stochastic fluctuations of the price series, capturing persistent economic forces beyond short-run noise \cite{nowotarski2016importance}. These components correspond naturally to distinct economic processes operating at different temporal scales: longer-horizon trends (e.g., semiannual and quarterly) reflect supplier procurement cycles, forward contract rollovers, and regulatory adjustments, while shorter-horizon components capture contract price updates and adjustments, promotional behavior, intra-firm competition, and high-frequency variation in posted retail prices. By constructing these components using only past information and feeding them into the causal graph discovery stage, we explicitly disentangle price formation mechanisms across horizons, stabilize estimation under nonstationarity, and enable economically interpretable attribution alongside accurate forecasts.

%% model framework

\begin{figure}[htbp]   
  \centering
  \includegraphics[width=1.15\linewidth]{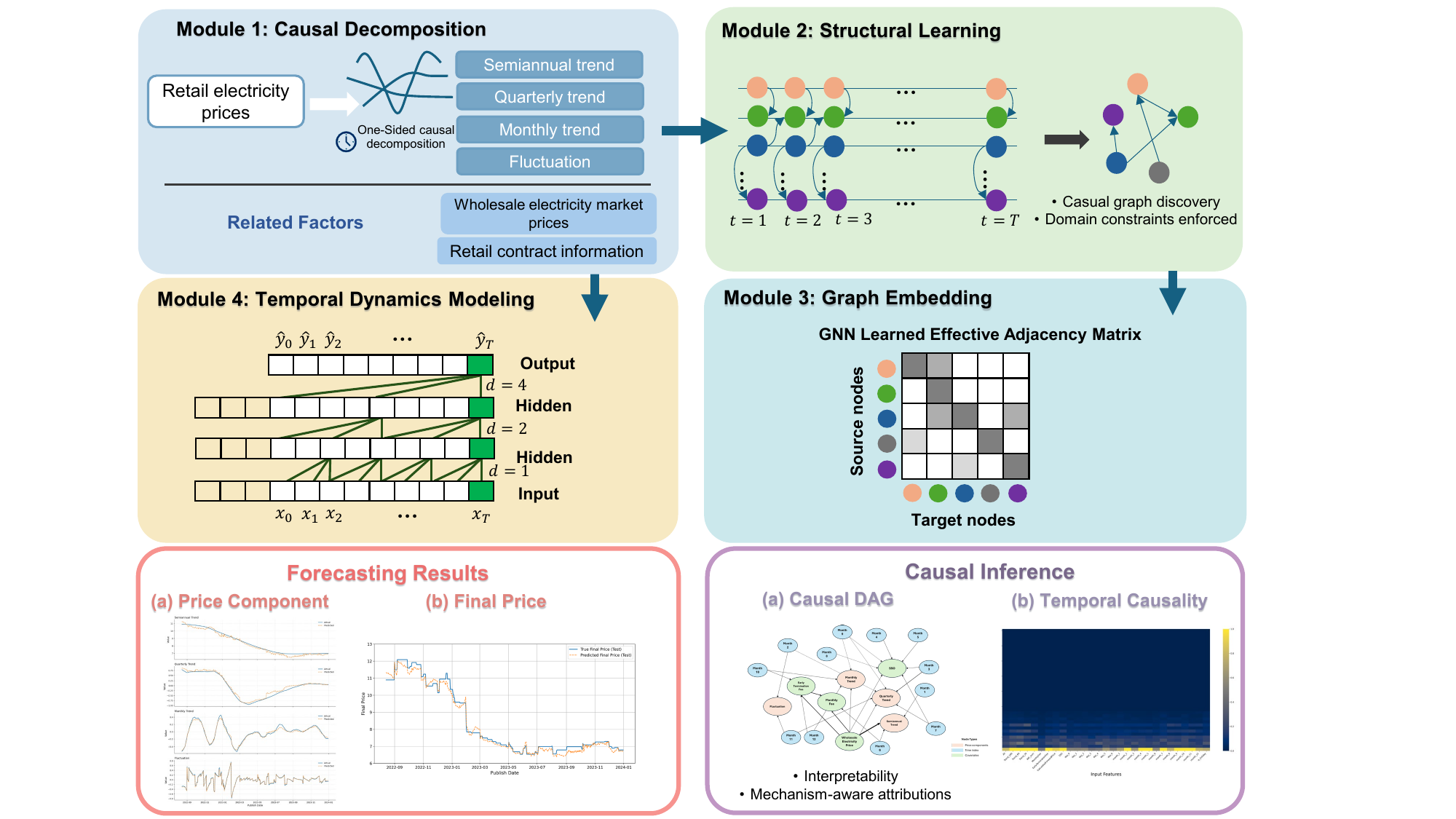}
  \caption{Overview of the proposed CG-TCN framework. Module 1: One-sided decomposition framework first decomposes retail electricity prices into multi-resolution components. Module 2 \& 3: NOTEARS then discovers a causal graph over these components and covariates, which is encoded as an adjacency embedding. Module 4: The learned causal structure guides the TCN to generate accurate forecasts and interpretable, mechanism-aware attributions.}
  \label{fig:Framework}
\end{figure}

Figure~\ref{fig:Framework} presents an overview of the proposed CG-TCN framework. In the first module, the retail electricity price series is decomposed into multiple temporal components—including semiannual, quarterly, and monthly trends, as well as short-term fluctuations. Next, a causal graph is discovered over these components and related market and contract features using the NOTEARS algorithm, capturing directional dependencies as a learned adjacency embedding. Finally, the causal embedding conditions %This causal embedding is then integrated into 
a TCN, which models temporal dynamics while respecting the causal structure. The result is a unified pipeline that couples %Overall, the framework unifies 
multi-scale decomposition, causal discovery, and neural forecasting to achieve accurate and interpretable, mechanism-aware price predictions. 

Specifically, the proposed framework contributes to the literature in four key ways:

%% innovations and contributions 
(1) We introduce a flexible one-sided causal multi-resolution decomposition framework based on STL and TSN that extracts interpretable temporal components of retail electricity prices in real time, capturing the dominant temporal dynamics that drive forecasting performance.

(2) We develop a causal discovery stage that identifies directional relationships between contract attributes, market fundamentals, and decomposed price components, and leverages these causal links to enhance predictive performance.

(3) We introduce a novel approach, CG-TCN, that combines the predictive strength of neural networks with the interpretability of causal inference. To our knowledge, this is the first approach that injects a learned causal DAG into a TCN through a graph-neural embedding, aligning convolutional representations with causal pathways %to unify causal structure and deep learning 
for applications in retail electricity price forecasting.

(4) We incorporate and tailor a spike-aware algorithm designed to improve the model’s forecasting performance during extreme price spike periods. The algorithm increases the representation of spike windows during training and applies a higher loss penalty to errors within these intervals, ensuring the model learns to better capture abrupt price surges.

(5) We demonstrate the proposed method on a comprehensive database from Ohio containing every daily retail choice offer filed by every retail supplier in every service territory for a decade. This demonstrates state-of-the-art forecasting accuracy and interpretable attributions across multiple horizons.

%% the paper's organization
The remainder of the paper is organized as follows. Section 2 introduces the proposed CG-TCN architecture and training. Section 3 describes data and variables. Section~4 reports the empirical results and forecasting performance. Section~5 discusses the main findings, interpretability, and policy implications. Section~6 concludes with a summary of contributions and directions for future research.

\section{The proposed CG-TCN forecasting framework}

The CG-TCN framework targets two complementary forms of causality that matter for retail electricity prices. The first is structural causality across variables: how wholesale energy forward prices, the standard service offer, and contract attributes propagate to multi-resolution price components. The second is temporal causality along the timeline: how past states influence future outcomes at multiple lags. Capturing only one of these leaves forecasts vulnerable to regime shifts. CG-TCN is built to encode both, explicitly and jointly.

The core methodological contribution is a DAG-conditioned temporal forecaster. We first estimate a sparse, directed causal graph over decomposed price components and covariates using a score-based discovery procedure with a differentiable acyclicity constraint. Domain rules are enforced so that targets do not act as parents and exogenous inputs remain exogenous, aligning the graph with economic directionality. This learned graph is then injected into the forecaster through a graph-neural embedding that aggregates information only along parent links and produces nodewise “causal features”. These causal features condition the temporal convolutions via feature fusion and channel-wise gating, so that signals arriving along plausible causal pathways receive higher gain during forecasting. The graph is fixed during training, but the conditioning is data-adaptive through learned weights.

To supply horizon-specific signals and reduce mode mixing, the observed retail price is decomposed into semiannual, quarterly, monthly, and high-frequency components, and these, together with market and contract covariates, enter both causal discovery and forecasting. A temporal convolutional backbone with dilated causal convolutions extracts long- and short-range lagged dependencies while preserving strict chronological order and using parameters efficiently at the available data scale. Fusing the graph-neural causal features with temporal features unifies structural and temporal causality in a single predictive representation.

Retail prices contain rare but consequential spikes. Training balances routine periods and stress periods through a spike-aware scheme that oversamples spike windows and increases their loss contribution while renormalizing batch weights for stability. The objective uses a robust pointwise discrepancy with component weights to emphasize accuracy on the most decision-relevant parts of the signal.

The framework yields decision-grade interpretability outputs: the learned causal graph that reflects cross-variable structure, the effective adjacency learned inside the forecaster that shows how the model actually routes information, and temporal saliency maps by target component and lag. Together, these interpretability outputs explain which drivers govern long-horizon trends versus short-term fluctuations and when those influences matter most, supporting procurement timing, hedging design, and policy evaluation in deregulated retail markets.

\begin{figure}[htbp]
  \centering
  \includegraphics[width=1\linewidth]{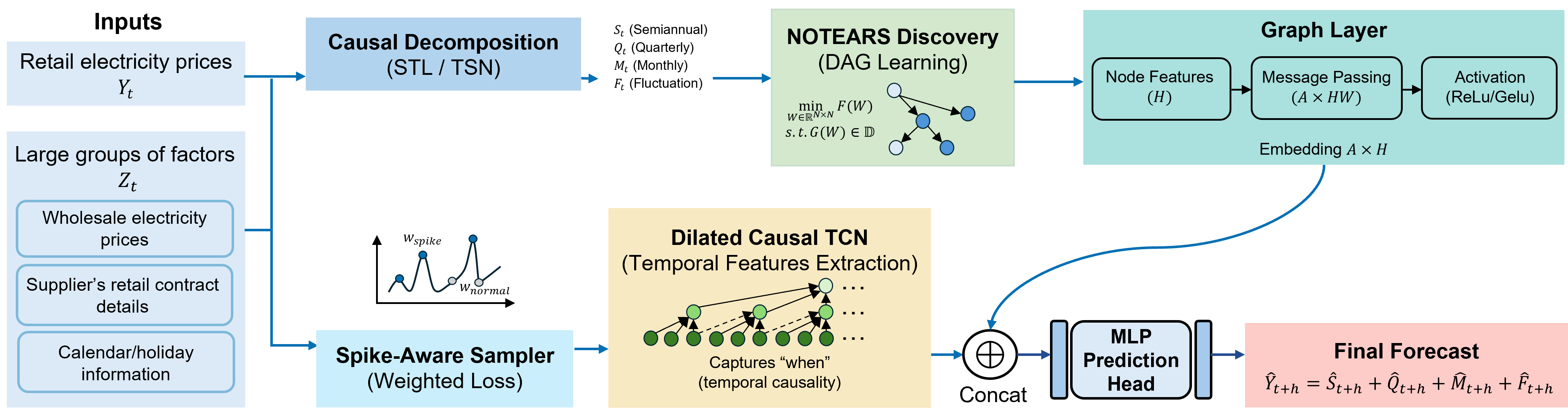}
  \caption{The framework of the proposed CG-TCN model.}
  \label{fig:model}
\end{figure}

Figure~\ref{fig:model} provides an overview of the proposed architecture, illustrating how each component contributes to interpretable and high-accuracy retail price forecasting.

The subsections that follow detail each stage in turn—multiscale decomposition of the price signal, domain-constrained causal graph discovery, graph-neural conditioning of the temporal convolutions, and the spike-aware training and interpretability outputs that complete the framework.

\subsection{Causal multi-scale time series decomposition}

Retail electricity markets in deregulated systems exhibit pronounced price volatility and complex interactions with wholesale forward and futures markets, resulting in nonstationary and nonlinear price behavior \cite{weron2014electricity}. These characteristics make it difficult for a single forecasting model to capture all underlying temporal patterns effectively. To address this challenge, we decompose the original price series into interpretable subcomponents at multiple temporal scales and use these as structured inputs for causal graph discovery and forecasting.

A key requirement in our application is \emph{real-time} implementability: at any time $t$, the decomposition must use only information available up to $t$ (i.e., be one-sided or causal), rather than relying on future observations \cite{hamilton2018you}. We therefore construct two alternative causal multi-scale decompositions of the daily retail price series, denoted $X_t$: (i) a nonparametric method based on a trailing-window adaptation of STL \cite{cleveland1990stl}, and (ii) a parametric method based on a bank of causal Butterworth low-pass filters, which we refer to as TSN \cite{percival2000wavelet, chen2025filter}. Both methods yield semiannual-, quarterly-, and monthly components, plus a high-frequency fluctuation term.

\subsubsection{Causal STL-based multi-scale decomposition}

Let $X_t$ denote the daily retail electricity price for a given utility region. 
For a given temporal scale with characteristic horizon $H$ (e.g., long-, medium-, or short-horizon), we define a trailing window as:
\begin{equation}
\mathcal{W}_t^{(H)} = \{X_{t-W_H+1}, \dots, X_t\},
\end{equation}
where $W_H$ is a window length proportional to $H$ (e.g., $W_H \approx 2H$). 
Applying the classical Seasonal--Trend decomposition using Loess (STL) to $\mathcal{W}_t^{(H)}$ yields:
\begin{equation}
X_\tau = T_\tau^{(H)} + S_\tau^{(H)} + R_\tau^{(H)}, \quad \tau \in \mathcal{W}_t^{(H)},
\end{equation}
where $T_\tau^{(H)}$ is a smooth trend component, $S_\tau^{(H)}$ is a seasonal component, and $R_\tau^{(H)}$ is the remainder. 
Because STL is applied only to past observations, the final trend estimate $T_t^{(H)}$ depends exclusively on $\{X_\tau : \tau \le t\}$ and can therefore be interpreted as a causal low-frequency estimate at time $t$.

To reduce computational cost and stabilize the estimated trend, we evaluate STL only at a sparse set of anchor times, defined as:
\begin{equation}
\mathcal{T}_s = \{t_k\}, \qquad t_{k+1} - t_k = s,
\end{equation}
where $s$ denotes a fixed stride (e.g., weekly). 
At each anchor time $t_k$, STL is applied to the trailing window $\mathcal{W}_{t_k}^{(H)}$, and the corresponding low-pass estimate is defined as:
\begin{equation}
L_{t_k}^{(H)} = T_{t_k}^{(H)}.
\end{equation}
For times between anchors, the full low-frequency series $\{L_t^{(H)}\}$ is obtained by linear interpolation between adjacent anchor values. 
This procedure yields a one-sided, real-time low-frequency component at scale $H$ while substantially reducing computational burden and preserving the interpretability of STL-based trend extraction.

In the empirical application, we consider three nested horizons $H_L > H_M > H_S$, corresponding approximately to semiannual, quarterly, and monthly dynamics. 
The associated low-pass trends $L_t^{(H_L)}$, $L_t^{(H_M)}$, and $L_t^{(H_S)}$ are combined into band-limited components via differencing:
\begin{align}
\text{Semiannual trend:} \quad & S_t^{\mathrm{STL}} = L_t^{(H_L)}, \\
\text{Quarterly component:} \quad & Q_t^{\mathrm{STL}} = L_t^{(H_M)} - L_t^{(H_L)}, \\
\text{Short-horizon component:} \quad & M_t^{\mathrm{STL}} = L_t^{(H_S)} - L_t^{(H_M)}, \\
\text{High-frequency fluctuation:} \quad & F_t^{\mathrm{STL}} = X_t - L_t^{(H_S)}.
\end{align}
This construction yields the additive decomposition:
\begin{equation}
X_t = S_t^{\mathrm{STL}} + Q_t^{\mathrm{STL}} + M_t^{\mathrm{STL}} + F_t^{\mathrm{STL}},
\end{equation}
where each component is updated in real time and depends only on information available up to time $t$. 
In the implementation, we set $(H_L, H_M, H_S)$ to correspond approximately to semiannual, quarterly, and monthly behavior, and use robust STL with linear trend and seasonal components.

\subsubsection{Causal TSN decomposition}

As a parametric benchmark, we implement an alternative causal multi-scale decomposition based on a bank of digital Butterworth low-pass filters. 
Let $\{X_t\}$ denote the daily retail electricity price series, sampled at frequency $f_s = 1$ cycle per day. 
For each cutoff horizon $H \in \{H_L, H_M, H_S\}$, we design a $K$th-order Butterworth low-pass filter with normalized cutoff frequency defined as:
\begin{equation}
\omega_H = \frac{1/H}{f_s/2},
\end{equation}
corresponding to attenuation of fluctuations with characteristic periods shorter than $H$. 
Applying the filter in causal (forward) form yields the following recursion:
\begin{equation}
B_t^{(H)} 
= \sum_{k=0}^K b_{H,k} \, X_{t-k}
- \sum_{k=1}^K a_{H,k} \, B_{t-k}^{(H)},
\end{equation}
where $\{a_{H,k}, b_{H,k}\}$ denote the filter coefficients. 
Because the recursion involves only current and past observations, the resulting low-pass component $B_t^{(H)}$ is strictly one-sided and implementable in real time.

Using the same nested horizons as in the STL-based approach, we form band-limited components by differencing successive low-pass outputs:
\begin{align}
\text{Semiannual trend:} \quad & S_t^{\mathrm{TSN}} = B_t^{(H_L)}, \\
\text{Quarterly component:} \quad & Q_t^{\mathrm{TSN}} = B_t^{(H_M)} - B_t^{(H_L)}, \\
\text{Short-horizon component:} \quad & M_t^{\mathrm{TSN}} = B_t^{(H_S)} - B_t^{(H_M)}, \\
\text{High-frequency fluctuation:} \quad & F_t^{\mathrm{TSN}} = X_t - B_t^{(H_S)}.
\end{align}
This yields the following additive decomposition:
\begin{equation}
X_t = S_t^{\mathrm{TSN}} + Q_t^{\mathrm{TSN}} + M_t^{\mathrm{TSN}} + F_t^{\mathrm{TSN}},
\end{equation}
which is linear and time-invariant. 
Unlike STL, this construction imposes an explicit frequency-domain structure and therefore provides a complementary, fully parametric perspective on multi-scale dynamics.

Together with the STL-based method, the TSN decomposition yields a collection of causal, horizon-specific components summarizing the retail electricity price series at semiannual, quarterly, monthly, and high-frequency scales. 
These representations are used as inputs for downstream causal graph discovery and forecasting. 

To assess the impact of decomposition choices, we evaluate several combinations of horizon lengths. Because these window lengths affect the smoothness, phase lag, and predictive usefulness of the resulting components, we compare one-step-ahead forecasting performance across all settings. The full results are reported in~\ref{sec:appendix_window}. The best-performing horizon settings are therefore used in all subsequent experiments.

\subsection{Causal graph discovery}

While the decomposition step effectively captures temporal heterogeneity, it does not explain why certain price movements occur or how different factors interact to generate them. To address this limitation, we incorporate a causal discovery stage that uncovers the structural dependencies among decomposed price components and exogenous covariates. Graph-based representations provide a natural framework for analyzing such complex systems: nodes denote variables, and directed edges represent potential causal influences. In particular, DAGs are well suited for modeling electricity market dynamics, as they encode directional and non-cyclic relationships consistent with economic causality (e.g., wholesale price shocks influencing retail contract adjustments). By learning the structure of a DAG directly from observational data, we can identify the underlying mechanistic pathways that drive retail price formation. This enables the model to move beyond correlation-based associations toward interpretable, causally grounded insights.

Formally, let $X \in \mathbb{R}^{T \times N}$ denote the data matrix consisting of $T$ observations of $N$-dimensional random vector $X = (X_1,\ldots,X_N)$. $X$ concatenates the multiscale price components (semiannual, quarterly, monthly, fluctuation) and covariates (e.g., forward price, SSO, contract fees, calendar dummies). Our objective is to learn a DAG $\mathcal{G} \in \mathbb{D}$, where $\mathbb{D}$ is the space of DAGs with $N$ nodes, that best models the joint distribution $\mathbb{P}(X)$ \cite{spirtes2000causation}. To achieve this, we adopt a structural equation model (SEM) of the form:
\begin{equation}
X_j = w_j^\top X + z_j, \qquad j=1,\dots,N,
\end{equation}
where $w_j$ is the weight vector encoding the influence of parent variables on $X_j$ and $z_j$ is an independent noise term. Let $W = [w_1 \cdots w_N] \in \mathbb{R}^{N \times N}$ denote the weighted adjacency matrix. The binary adjacency matrix $A(W) \in \{0,1\}^{N \times N}$ is then defined as:
\begin{equation}
[A(W)]_{ij} = 1 \Leftrightarrow w_{ij} \neq 0,
\end{equation}
which induces a directed graph $G(W)$.

The goal is to estimate $W$ such that the resulting graph $G(W)$ lies in the DAG space $\mathbb{D}$. This can be posed as the following optimization problem:
\begin{equation}
\min_{W \in \mathbb{R}^{N \times N}} F(W) \quad \text{s.t.} \quad G(W) \in \mathbb{D},
\label{eq:causal_opt}
\end{equation}
where $F(W)$ is a score function defined by a least-squares loss and an $\ell_1$-regularization term:
\begin{equation}
F(W) = \ell(W;X) + \lambda \|W\|_1
      = \frac{1}{2N}\|X - XW\|_F^2 + \lambda \|W\|_1.
\end{equation}
Here, $\ell(W;X)$ denotes the least-squares loss, $\|\cdot\|_F$ is the Frobenius norm, and $\|\cdot\|_1$ encourages sparsity of the adjacency matrix. 

The acyclicity constraint $G(W) \in \mathbb{D}$ is enforced through a smooth continuous characterization, the NOTEARS algorithm \cite{zheng2018dags}, allowing the problem to be solved using gradient-based optimization rather than discrete search. To align the learned structure with economic directionality and reduce spurious links, we impose a fixed edge mask $\mathcal{M}$ that prunes prohibited relationships during optimization: (i) forecasting targets are not permitted as parents; (ii) exogenous inputs (forward price, SSO, calendar dummies) are sources (no incoming edges); and (iii) redundant edges among purely time-index indicators are disallowed. The NOTEARS algorithm iteratively updates the weighted adjacency matrix $W$ by computing the gradient of the objective function, applying a soft-thresholding operator to promote sparsity, and ensuring acyclicity through a differentiable penalty term. Convergence is determined when changes in either $W$ or the objective function fall below a tolerance threshold. After convergence, a post-processing step is applied to prune weak edges by thresholding small weights to zero. This continuous optimization framework enables NOTEARS to efficiently learn DAG structures at scale, achieving accurate causal discovery without resorting to combinatorial search procedures. The resulting weighted, sparse adjacency serves as a causal prior that will be encoded in the forecasting model, enabling the downstream network to privilege information flow along discovered structural pathways.

\subsection{Graph neural network with causal embedding}

To capture the structural dependencies revealed by the causal graph, we incorporate a GNN layer that explicitly propagates information along the edges of the learned adjacency matrix \cite{kipf2016semi,wu2020comprehensive}. This design allows the temporal forecasting model not only to exploit time-series dynamics, but also to embed domain-specific causal relationships between variables- for example, allowing wholesale electricity prices to influence retail price components while preventing information flow in the reverse direction. This layer is the vehicle for our DAG-to-forecaster infusion: a graph learned in the discovery stage is converted into a direction-aware embedding that later conditions the temporal convolutions, thereby coupling structural and temporal causality inside a single predictor.

Let $X \in \mathbb{R}^{B \times L \times N}$ denote a batch of input 
time series with batch size $B$, sequence length $L$, and $N$ variables. From the causal discovery stage, we obtain a weighted adjacency matrix 
$A \in \mathbb{R}^{N \times N}$, where each entry is defined as:
\begin{equation}
    A_{ij} = W_{\text{causal}, i, j},
\end{equation}
representing the causal influence of variable $j$ on variable $i$. 
To ensure that each node aggregates information evenly from its parents, 
we normalize the adjacency matrix as:
\begin{equation}
    \hat{A}_{\text{in}} = D_{\text{in}}^{-1} A,
\end{equation}
where $D_{\text{in}}$ is the in-degree diagonal matrix. This normalization 
ensures that the update for each node corresponds to an average over its 
causal parents. During training and evaluation, a fixed edge mask consistent with the discovery-stage domain rules is applied to $A$ (and thus to $\hat{A}_{\text{in}}$), guaranteeing that prohibited relations remain inactive and that information propagates only along economically admissible directions.

Given the normalized adjacency $\hat{A}_{\text{in}}$ and hidden features $H$, the GNN layer first computes an intermediate 
update:
\begin{equation}
    H^{+} = \alpha \, \hat{A}_{\text{in}} H W_{\text{in}}
           + \beta \, H W_{\text{self}} + b,
\end{equation}
where $W_{\text{in}}$ and $W_{\text{self}}$ are learnable weight matrices 
for parent-aggregated and self-updates, respectively, $b$ is a learnable 
bias, and $\alpha, \beta > 0$ are scalars learned during training that balance the contribution of parent and self-features. The final node representation is then obtained by applying a residual connection, nonlinearity, and normalization:
\begin{equation}
    H_{\text{new}} = \sigma(H^{+} + H),
\end{equation}
where $\sigma(\cdot)$ is a leaky-ReLU activation and normalization is applied for stability.

To form a compact representation, we aggregate node-level embeddings into 
a causal embedding by averaging across variables:
\begin{equation}
    C_t = \frac{1}{d} \sum_{i=1}^d H_{\text{new}}[i,:],
\end{equation}
where $d$ is the number of variables (nodes) and $H_{\text{new}}[i,:]$ 
denotes the embedding of node $i$. The resulting 
$C_t \in \mathbb{R}^{B \times L \times d_{\text{gnn}}}$ summarizes how 
each variable is influenced by its causal parents at time $t$, with $d_{\text{gnn}}$ denoting the output feature dimension of the GNN layer. Because $H$ is computed from the current temporal context while the graph is fixed within a fold, $C_t$ is time-varying and reflects context-dependent amplification of parent signals consistent with the learned structure.
%It provides a structured embedding that complements temporal features in the forecasting model.

This causal embedding serves two critical roles in the overall framework. It injects the discovered causal structure into the latent representation, ensuring that information flow within the model respects domain-relevant directionality rather than arbitrary correlations. While the TCN captures sequential dependencies, the causal GNN captures cross-variable dependencies, producing an embedding that bridges structural and temporal learning. In the next subsection, this embedding is fused with temporal features and used to gate convolutional channels, so that features aligned with discovered structural pathways are given higher effective gain during forecasting. Together, these mechanisms allow the CG-TCN to learn representations that are not only predictive but also interpretable in terms of causal mechanisms, revealing how wholesale electricity market prices or retail contract information propagate through the price formation process.

\subsection{Temporal convolutional network and feature fusion}

The temporal convolutional network (TCN) is responsible for capturing sequential dependencies in the input series \cite{lea2017temporal,bai2018empirical}. Unlike standard convolutions, the TCN employs dilated causal convolutions, which expand the receptive field without increasing the number of parameters and ensure that predictions at time $t$ depend only on information available up to $t-1$. The TCN consists of a stack of $L_{\text{tcn}}$ one-dimensional causal convolutional blocks with exponentially increasing dilation factors and residual connections. Each block applies convolution, nonlinearity, normalization, and dropout, yielding an effective receptive field that spans the longest decomposition horizon (semiannual) while maintaining sample efficiency.

Formally, given an input sequence $X_{1:t-1}$, the TCN produces a sequence of temporal feature representations:
\begin{equation}
T_{1:t-1} = f_{\text{tcn}}(X_{1:t-1}),
\end{equation}
where $f_{\text{tcn}}(\cdot)$ denotes the stacked dilated causal convolutional operations and
$T_{1:t-1} \in \mathbb{R}^{(t-1) \times d_{\text{tcn}}}$. We use the feature vector at the most recent time step, $T_t \in \mathbb{R}^{d_{\text{tcn}}}$, as a summary of historical temporal patterns.

In parallel, a graph neural network (GNN) processes the learned causal adjacency matrix to produce a causal embedding $C_t \in \mathbb{R}^{d_{\text{gnn}}}$ that summarizes structural dependencies among decomposed price components and exogenous drivers at time $t$. To integrate temporal and structural information, we fuse these representations through feature concatenation:
\begin{equation}
Z_t = \text{Concatenate}(T_t, C_t),
\end{equation}
where $Z_t \in \mathbb{R}^{d_{\text{tcn}} + d_{\text{gnn}}}$ combines temporal dynamics (“what happened”) with structural context (“why it happened”).

The fused representation $Z_t$ is then passed to a forecasting head implemented as a two-layer multilayer perceptron (MLP) with a nonlinear activation and dropout regularization. This head projects the fused features onto the output space to produce multi-step forecasts of the decomposed price components:
\begin{equation}
\hat{Y}_{t+1:t+H} = f_{\text{out}}(Z_t),
\end{equation}
where $\hat{Y}_{t+1:t+H} \in \mathbb{R}^{H \times 4}$ contains the predicted semiannual, quarterly, monthly, and high-frequency fluctuation components over horizon $H$.

Finally, the overall retail electricity price forecast is reconstructed by summing the predicted components at each horizon:
\begin{equation}
\hat{P}_{t+h} =
\hat{Y}_{t+h}^{(\text{semiannual})} +
\hat{Y}_{t+h}^{(\text{quarterly})} +
\hat{Y}_{t+h}^{(\text{monthly})} +
\hat{Y}_{t+h}^{(\text{fluctuation})},
\qquad h = 1,\dots,H.
\end{equation}

This fusion strategy allows the model to jointly leverage temporal persistence captured by the TCN and structural dependencies encoded by the causal graph, without imposing additional parametric constraints on their interaction. Ablation results in Section~4.6 confirm that incorporating causal embeddings alongside temporal features substantially improves forecasting accuracy and stability compared to purely temporal baselines. Because the causal embedding encodes structural dependencies among decomposed price components and exogenous drivers, the component-level outputs yield actionable attributions—for example, wholesale market signals dominating long-horizon components and contract attributes shaping high-frequency behavior—thereby clarifying the mechanisms underlying each forecasted price movement.

\subsection{Weighed loss function}

Forecasting retail electricity prices is particularly challenging because forecast errors over time are highly asymmetric: most days exhibit relatively stable price patterns with small errors, while infrequent spike events generate disproportionately large errors and contribute outsized system risk and consumer costs. Standard loss functions that weight all samples equally tend to bias models toward fitting the majority of stable observations, at the expense of accuracy during these rare but consequential spike events. To address this imbalance, we design a weighted loss function that jointly accounts for multi-resolution price components and explicitly emphasizes spike windows during training.

Our objective couples a component-weighted, sample-weighted loss with a robust pointwise discrepancy. Let $\hat{Y}$ denote the model prediction and $Y$ the ground truth. Each target is decomposed into $C=4$ components (semiannual trend, quarterly trend, monthly trend, fluctuation). For single-step forecasting,
let $\hat{y}_{i,c}$ and $y_{i,c}$ be the prediction and label of sample $i$ for component $c\in\{1,\dots,4\}$. For multi-step forecasting with horizon $H$, we index the step by $t\in\{1,\dots,H\}$ and write $\hat{y}_{i,t,c}$.

We use the Smooth~L1 discrepancy with unit threshold \cite{huber1992robust},
\begin{equation}
\ell_{\text{smoothL1}}(a,b)=
\begin{cases}
\frac{1}{2}(a-b)^2, & |a-b|<1,\\[2pt]
|a-b|-\frac{1}{2}, & \text{otherwise}.
\end{cases}
\label{eq:huber}
\end{equation}
This formulation is less sensitive to outliers than mean squared error (MSE) and more stable than mean absolute error (MAE). For small errors, the loss is quadratic, behaving like MSE and providing smooth gradients for stable convergence. For large errors, the loss is linear, behaving like MAE and making the model more robust to outliers such as extreme price spikes.

Let $\boldsymbol{\alpha}\!=\!(\alpha_1,\ldots,\alpha_C)^\top\!>\!0$ be learnable or user-set component weights. For single-step training, the per-sample component-weighted loss is defined as:
\begin{equation}
\tilde{\ell}_i
= \sum_{c=1}^{C} \alpha_c \, \ell_{\text{smoothL1}}\!\left(\hat{y}_{i,c},\, y_{i,c}\right).
\label{eq:comp-weight}
\end{equation}
For multi-step training, we average across the horizon:
\begin{equation}
\tilde{\ell}_i
= \frac{1}{H}\sum_{t=1}^{H}\sum_{c=1}^{C} \alpha_c \,
\ell_{\text{smoothL1}}\!\left(\hat{y}_{i,t,c},\, y_{i,t,c}\right).
\label{eq:multi-h}
\end{equation}

% Forecasting retail electricity prices is particularly challenging because forecast errors over time are highly asymmetric: most days exhibit relatively stable price patterns with small errors, while infrequent spike events generate disproportionately large errors and contribute outsized system risk and consumer costs. Standard loss functions, which treat all samples equally, tend to bias the model toward minimizing errors on the majority of stable observations, at the expense of accuracy on these rare but critical spikes. This issue has been well documented in electricity price forecasting, where spikes are recognized as key drivers of volatility and financial risk \cite{uniejewski2019importance, nygaard2025enhancing}. To address this imbalance, we incorporate and tailor a sample-weighting scheme that explicitly amplifies the contribution of spike windows during training. Implementation details of the spike-aware algorithm, including spike detection criteria, window construction, and diagnostics on spike frequency, are reported in~\ref{sec:appendix_spike}.

To further address the imbalance induced by rare but impactful price spikes, we incorporate a sample-weighting scheme that explicitly amplifies the contribution of spike windows during training. This issue has been widely documented in electricity price forecasting, where spikes are recognized as key drivers of financial risk and forecasting error \cite{uniejewski2019importance, nygaard2025enhancing}. Implementation details of the spike-aware algorithm, including spike detection criteria, window construction, and diagnostics on spike frequency, are reported in Section~\ref{sec:appendix_spike}.

Specifically, each sample $i$ receives a weight $w_i > 0$. Inside each batch, we 
normalize these weights to ensure scale-invariance:
\begin{equation}
\tilde{w}_i \;=\; \frac{w_i}{\frac{1}{B}\sum_{j=1}^{B} w_j},
\qquad
\text{so that } \;\; \frac{1}{B}\sum_{i=1}^{B} \tilde{w}_i \,=\, 1,
\label{eq:weight-norm}
\end{equation}
where $B$ is the batch size. The final batch loss is the weighted mean of
per-sample losses:
\begin{equation}
\mathcal{L}_{\text{batch}}
= \frac{1}{B}\sum_{i=1}^{B} \tilde{w}_i \, \tilde{\ell}_i.
\label{eq:batch-loss}
\end{equation}

This weighted loss design balances three objectives: (i) robustness, since the Smooth~L1 loss adapts between MSE-like and MAE-like behavior; (ii) interpretability, by allowing component-specific weights ${\alpha}$ to emphasize different decomposition modes; and (iii) event emphasis, by assigning higher weights to spike windows without destabilizing training through batch-wise normalization. Together, these elements yield more stable optimization and improved predictive accuracy, especially in the presence of volatility and price spikes.

\section{Data}

% We evaluate the proposed CG-TCN framework using a comprehensive database containing every residential retail electricity contract filed by every Competitive Retail Electric Service (CRES) supplier/marketer in the State of Ohio for approximately one decade. From the perspective of residential energy affordability and social welfare, contract prices (cents per kWh) are a primary outcome of interest. In addition to prices, we incorporate detailed contract-level attributes, including monthly fees, early termination fees, and other retail design features, as well as market-level information such as forward wholesale electricity prices and default utility service rates. Integrating these dimensions allows the model to jointly capture pricing behavior, contractual heterogeneity, and broader market conditions, thereby enhancing both predictive accuracy and economic interpretability.

We evaluate the proposed CG-TCN framework using a newly constructed, comprehensive database of residential retail electricity contracts from Ohio’s deregulated electricity market. All competitive retail electricity supply offers in Ohio are required to be filed by registered Competitive Retail Electric Service (CRES) suppliers with the Public Utilities Commission of Ohio (PUCO) through an official regulatory web portal and are made publicly available to consumers via a daily retail choice marketplace. Leveraging this regulatory requirement, we built a contract-level relational database containing the full universe of residential retail electricity offers filed with PUCO between January 2014 and December 2023. Unlike commonly used aggregated datasets (e.g., utility-level or EIA summaries), these data capture consumer-facing retail prices and contractual terms as they are posted in real time, enabling a detailed analysis of retail price formation and forecasting at a level of granularity that is rarely available in the literature.

From the perspective of residential energy affordability and social welfare, contract prices (cents per kWh) are the primary outcome of interest. In addition to posted prices, the database includes rich contract-level attributes such as fixed monthly fees, early termination fees, contract duration, rate structure, and renewable content where applicable, along with other retail design features. We further integrate market-level information, including wholesale forward electricity prices and default utility service rates. Combining these dimensions allows the model to jointly capture supplier pricing behavior, contractual heterogeneity, and broader market conditions, thereby enhancing both predictive accuracy and economic interpretability. 

Ohio operates one of the largest and most robust deregulated retail electricity markets in the United States. Its retail choice operations are expansive, on par with or exceeding states like Texas and Pennsylvania. Over the past decade, approximately one hundred residential electricity offers with heterogeneous contract terms have been published daily by more than twenty competitive suppliers within each of the state’s six electric distribution utility service territories. In this study, we focus on the Duke Energy Ohio service territory. This is the service area represented by the greater Cincinnati metro area and outlying regions. 
%Prior work has indicated that retail supply offers in the Duke territory tend to exhibit particularly pronounced pricing heterogeneity and strategic supplier behavior, making it a good testbed for the methodology introduced here. It also provides a representative and policy-relevant setting for studying retail price formation \cite{dormady2025efficiency, ji2025renewable}. 
Prior work has indicated that retail supply offers in the Duke territory exhibit particularly pronounced pricing heterogeneity and strategic supplier behavior, making it a well-suited testbed for evaluating the proposed methodology. At the same time, the Duke territory provides a representative and policy-relevant empirical setting for studying retail price formation \cite{dormady2025efficiency, ji2025renewable}.
Using Duke Energy Ohio data, we construct two daily price series for 12-month fixed residential contracts over the period from January 21, 2014, to December 31, 2023: the daily median price and the daily minimum price. The median price reflects the central tendency of available offers on a given day and serves as a representative measure of typical market pricing faced by residential consumers. In contrast, the daily minimum price captures the lowest available contract price observed each day, representing the best-case option accessible to price-sensitive consumers and providing insight into the lower bound of market competition.

\begin{figure}[htbp]
  \centering
  \includegraphics[width=0.8\linewidth]{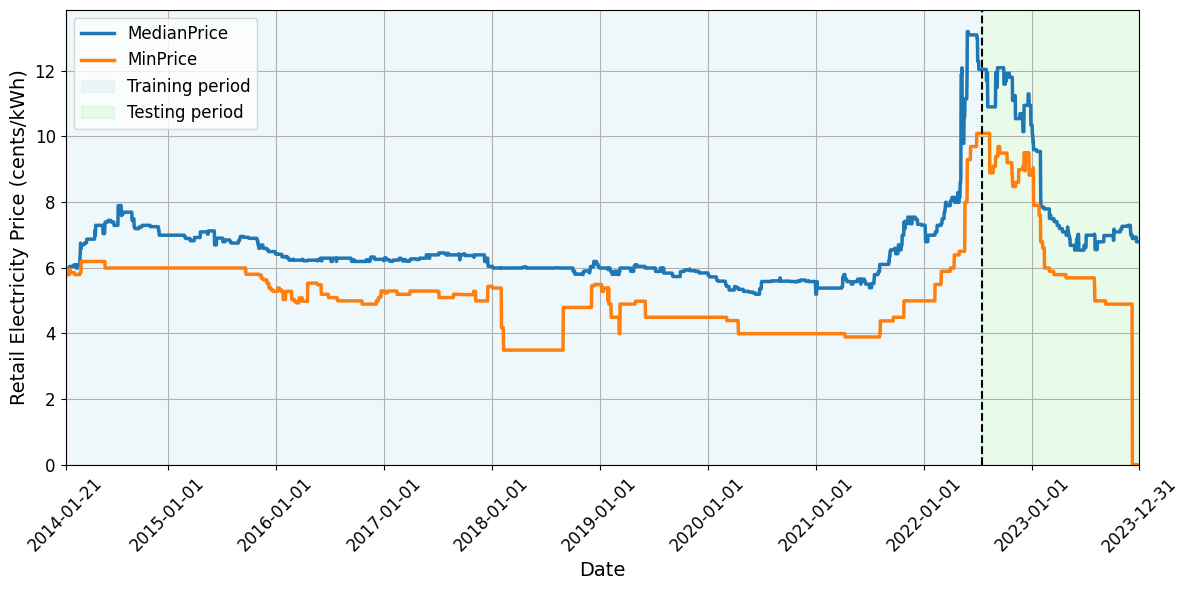}
  \caption{Time series of daily median and minimum retail electricity prices.}
  \label{fig:time_series}
\end{figure}

Unless otherwise stated, the empirical analysis and results discussion primarily focus on the median price series, as it offers a representative and policy-relevant benchmark for assessing overall market dynamics and forecasting performance. The minimum price series is analyzed in parallel to evaluate model robustness under more volatile and extreme pricing conditions. Both series are independently divided into training (85\%) and testing (15\%) subsets. To preserve temporal ordering, time-series cross-validation is applied within the training period for model selection, and final model performance is evaluated on the held-out test set. Table~\ref{tab:train_test_split} summarizes the sample periods and data splits, while Figure~\ref{fig:time_series} and Table~\ref{tab:price_stats} present the corresponding time-series trajectories and descriptive statistics. Details of the time-series cross-validation procedure are provided in~\ref{sec:appendix_tscv}.

\begin{table}[htbp]
\centering
\caption{Description of daily aggregated retail electricity price series.}
\scriptsize   
\begin{tabular}{llccc}
\hline
\textbf{Dataset} & \textbf{Subset} & \textbf{Period} & \textbf{\# Days (aggregated)} & \textbf{Proportion} \\
\hline
\multirow{3}{*}{MinPrice} 
 & Training set & 2014-01-21 -- 2022-07-15 & 3088 & 85\% \\
 & Test set     & 2022-07-16 -- 2023-12-31 & 534  & 15\% \\
 & All          & 2014-01-21 -- 2023-12-31 & 3622 & 100\% \\
\hline
\multirow{3}{*}{MedianPrice} 
 & Training set & 2014-01-21 -- 2022-07-15 & 3088 & 85\% \\
 & Test set     & 2022-07-16 -- 2023-12-31 & 534  & 15\% \\
 & All          & 2014-01-21 -- 2023-12-31 & 3622 & 100\% \\
\hline
\end{tabular}

\vspace{2pt}
\footnotesize\textit{Note:} Each daily observation corresponds to a median or minimum price computed from multiple underlying residential retail electricity contract offers. The full dataset contains over two million individual contract observations.
\label{tab:train_test_split}
\end{table}

\begin{table}[htbp]
\centering
\caption{Statistical information of the daily retail electricity prices.}
\resizebox{\textwidth}{!}{%
\begin{tabular}{llrrrrrrrr}
\hline
\textbf{Dataset} & \textbf{Subset} & \textbf{Mean} & \textbf{Standard deviation} & \textbf{Minimum} & \textbf{Median} & \textbf{Maximum} & \textbf{Range} & \textbf{Skewness} & \textbf{Kurtosis} \\
\hline
\multirow{3}{*}{MinPrice} 
 & Training set & 5.0407 & 1.0035 & 3.4900 & 4.9900 & 10.0900 & 6.6000 & 1.4879 & 5.8515 \\
 & Test set     & 6.7210 & 2.1125 & 0.0000 & 5.7900 & 10.0900 & 10.0900 & -0.2667 & 0.4037 \\
 & All          & 5.2643 & 1.3953 & 0.0000 & 5.0900 & 10.0900 & 10.0900 & 1.1261 & 3.8409 \\
\hline
\multirow{3}{*}{MedianPrice} 
 & Training set & 6.4306 & 1.0942 & 5.1900 & 6.2300 & 13.1900 & 8.0000 & 3.7000 & 18.4269 \\
 & Test set     & 8.5989 & 2.0257 & 6.5350 & 7.2900 & 12.0900 & 5.5550 & 0.6055 & -1.3759 \\
 & All          & 6.7438 & 1.4795 & 5.1900 & 6.2900 & 13.1900 & 8.0000 & 2.4910 & 6.3449 \\
\hline
\end{tabular}%
}

\vspace{2pt}
\footnotesize\textit{Note:} Price statistics are reported in cents per kilowatt-hour (¢/kWh). Each daily observation corresponds to an aggregated retail price (median or minimum) computed from multiple underlying residential contract offers.
\label{tab:price_stats}
\end{table}

Building on this framework, we operationalize the covariates using two groups of price-relevant factors: a) retail suppliers' individual contract attributes, and b) market conditions. These are also combined with calendar-based time features, as summarized in~\ref{sec:appendix_variables}. Contract attributes include the fixed monthly fee (USD) and the early termination fee (USD). Market variables comprise two key explanatory features. First, we use the power forward (or futures) price that clears on the very same day as each contract's date of origination. These data come from S\&P Market Intelligence, and we utilize power forwards for PJM's West Hub, which is the proximate power trading hub for the wholesale market for the Duke service territory and the most liquid trading hub in PJM. The wholesale forward strips are aligned to the 12-month delivery horizon of each day’s offers and converted from dollars per megawatt hour to cents per kWh (see~\ref{sec:appendix_forward} for strip definitions, averaging rules, and unit conversions). Additionally, calendar indicators capture day-of-week effects, monthly seasonality, and U.S. federal holidays. All covariates are aligned at a daily frequency and merged with the corresponding price aggregates.

\section{Results and Discussions}

\subsection{Experimental design}

\subsubsection{Evaluation metrics}
To evaluate forecasting performance, we use a set of widely adopted error and fit metrics, including mean absolute error (MAE), mean squared error (MSE), root mean squared error (RMSE), mean absolute percentage error (MAPE), normalized root mean squared error (NRMSE), and $R^2$, which are defined as: 

% \begin{equation}
% ND = \frac{\sum_{t=1}^{T} |y_t - \hat{y}_t|}{\sum_{t=1}^{T} |y_t|}, 
% \label{eq:nd} 
% \end{equation}

\begin{equation}
MAE = \frac{1}{T} \sum_{t=1}^{T} |y_t - \hat{y}_t|, 
\label{eq:mae} 
\end{equation}

\begin{equation}
MSE = \frac{1}{n} \sum_{t=1}^{n} (y_t - \hat{y}_t)^2, 
\label{eq:mse}
\end{equation}

\begin{equation}
RMSE = \sqrt{MSE}, 
\label{eq:rmse}
\end{equation}

\begin{equation}
MAPE = \frac{100\%}{T} \sum_{i=1}^{T} \left| \frac{y_i - \hat{y}_i}{y_i} \right|, \label{eq:mape}
\end{equation}

\begin{equation}
NRMSE = \frac{100\% \times RMSE}{y_{\max} - y_{\min}}, 
\label{eq:nrmse}
\end{equation}

\begin{equation}
R^2 = 1 - \frac{\sum_{t=1}^{T} (y_t - \hat{y}_t)^2}{\sum_{t=1}^{T} (y_t - \bar{y})^2}. \label{eq:r2}
\end{equation}
where $\hat{y}_t$ is the predicted value of the actual value $y_t$, and $T$ is the time scope of the test set. The model performs better when MAE, MSE, RMSE, MAPE, and NRMSE are lower, and $R^2$ is higher.

\subsubsection{Baseline methods}

For comparison, we introduce several candidate models to evaluate the performance of the proposed CG-TCN.

  \textbf{(1) RF:} Random Forest is an ensemble learning method that constructs multiple decision trees and aggregates their outputs to improve predictive accuracy. For time series forecasting, it requires reframing the sequential data as a supervised learning task \cite{breiman2001random}.
  
  \textbf{(2) XGB:} XGBoost (eXtreme Gradient Boosting) is a scalable and efficient implementation of gradient boosting. It builds trees sequentially, where each tree corrects the errors of the previous ones, resulting in strong predictive performance and robustness \cite{chen2016xgboost}.
  
  \textbf{(3) AR:} The Autoregressive (AR) model predicts the current value of a time series using a linear combination of its past values, making it a simple yet effective baseline for temporal dependence \cite{box2015time}.
  
  \textbf{(4) LSTM:} Long Short-Term Memory (LSTM) is a recurrent neural network architecture designed to capture both short- and long-term dependencies in sequential data. It employs memory cells with input, output, and forget gates to regulate the flow of information, enabling accurate time series forecasting \cite{hochreiter1997long}.
  
  \textbf{(5) TCN:} The Temporal Convolutional Network (TCN) is a deep learning architecture for sequential data that applies causal and dilated convolutions to capture long-range dependencies. By preserving the temporal order and enabling parallel computation, TCNs provide efficient and accurate time series forecasting \cite{lin2021temporal}. This baseline receives the same inputs as CG-TCN but without any causal graph conditioning.
  
  \textbf{(6) MTGNN:} The Multivariate Time Series Graph Neural Network (MTGNN) integrates graph learning, graph convolution, and temporal convolution to jointly model spatial and temporal dependencies in multivariate time series \cite{wu2020connecting}. We use its built-in graph learning; no external DAG is injected, ensuring a clean contrast with our DAG-conditioned approach.
  
  \textbf{(7) Autoformer:} Autoformer is a Transformer-based forecasting model that introduces a decomposition architecture and an Auto-Correlation mechanism, enabling it to capture both seasonal-trend components and long-range temporal dependencies effectively \cite{wu2021autoformer}. Encoder–decoder depths and moving-average factors in this model are tuned via Optuna.
  
  \textbf{(8) CTCN:} The Causal Temporal Convolutional Network (CTCN) extends the Temporal Convolutional Network by embedding causal inference into its structure. Using dilated causal convolutions, it ensures predictions rely only on past information, while adaptive group LASSO penalties on input weights identify nonlinear Granger causal relationships among variables. This design enables CTCN to achieve both accurate time series forecasting and interpretable causal discovery \cite{li2023forecasting}. CTCN receives the same inputs but does not use a pre-learned, domain-constrained DAG; this serves as the closest causal baseline to isolate the incremental value of our DAG-to-TCN conditioning.

\subsection{Time series decomposition results}

An appropriate decomposition of retail electricity price series is essential for forecasting applications, as prices exhibit strong nonstationarity, nonlinear dynamics, and scale-dependent variability. To address these features, we apply causal multi-scale time series decomposition directly to the observed price data, separating long-run trends from intermediate cycles and short-term fluctuations.

\begin{figure}[htbp]
  \centering
  \includegraphics[width=0.8\linewidth]{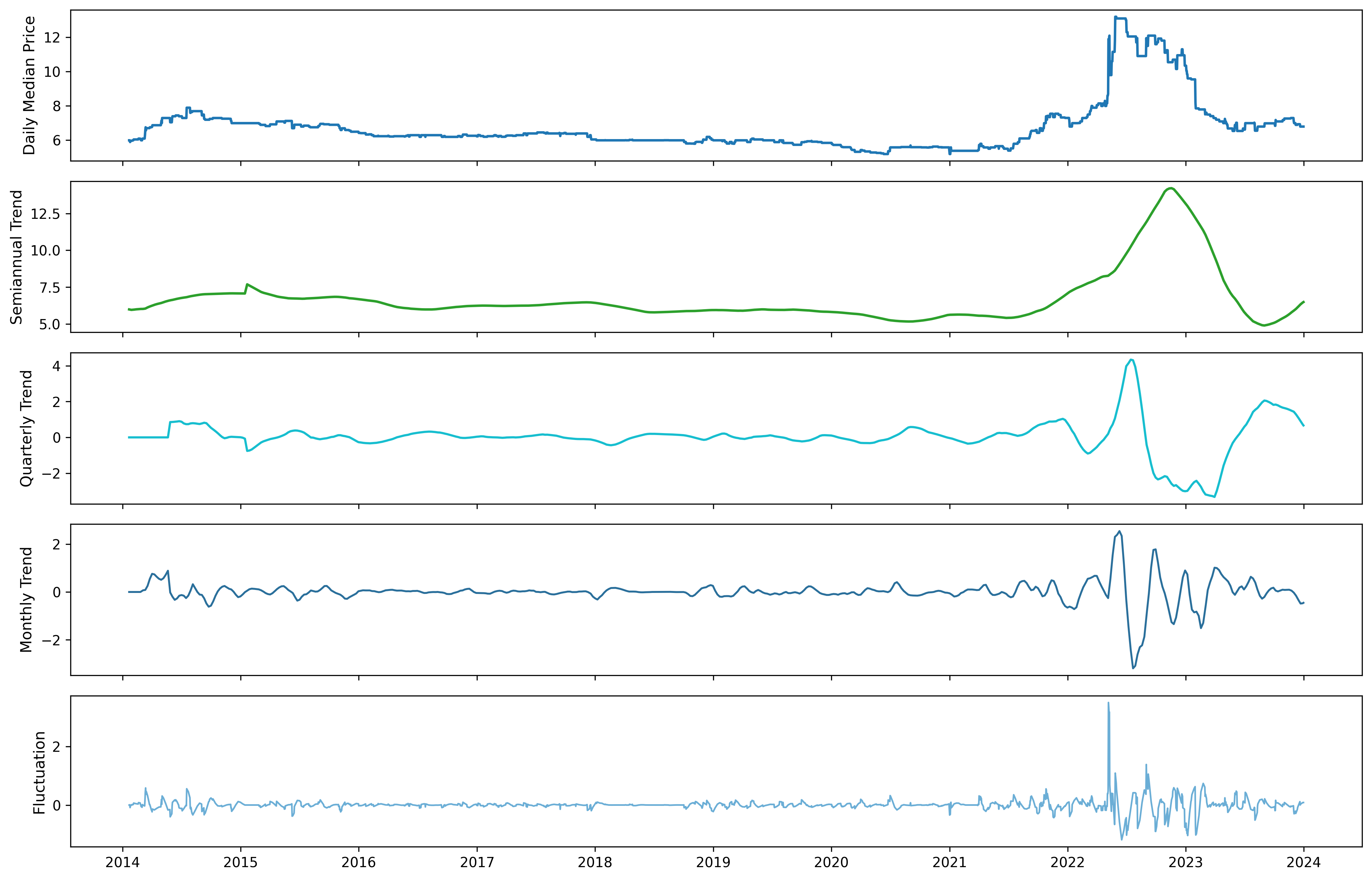}
  \caption{Causal multi-scale decomposition of retail electricity daily MEDIAN prices using STL (180-60-20 windows).}
  \label{fig:decomposition_plot_median}
\end{figure}

Figure~\ref{fig:decomposition_plot_median} presents the causal STL-based decomposition of daily retail electricity \emph{median} prices, using trailing windows of 180 days for the semiannual trend, 60 days for the quarterly component, and 20 days for the monthly component. The extracted semiannual trend captures gradual market-wide price movements, while the quarterly and monthly components reflect evolving seasonal patterns and transient deviations. The high-frequency fluctuation term isolates short-lived shocks and noise, demonstrating the ability of the STL-based approach to flexibly represent nonlinear dynamics in median retail prices.

\begin{figure}[htbp]
  \centering
  \includegraphics[width=0.8\linewidth]{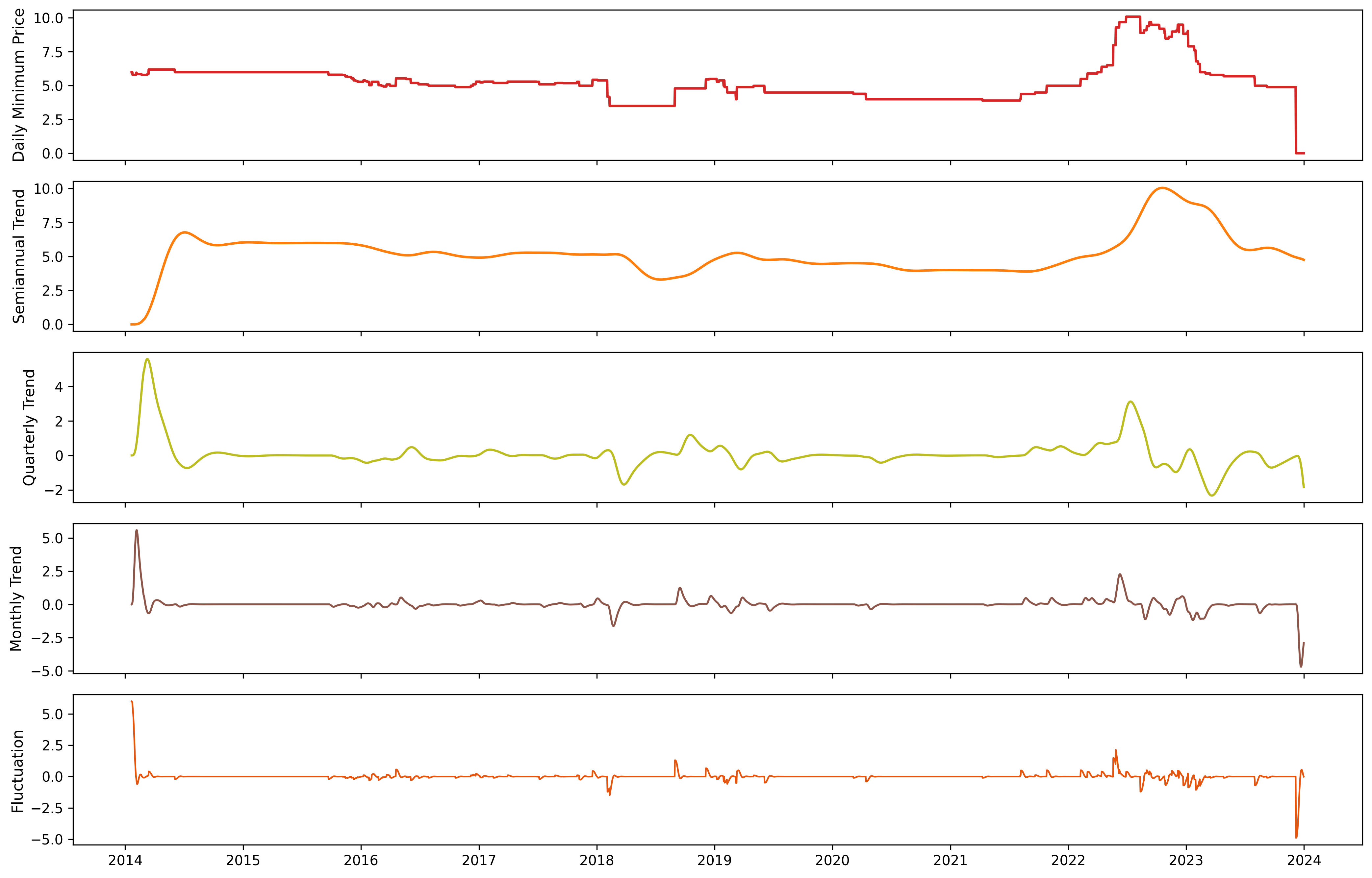}
  \caption{Causal multi-scale decomposition of retail electricity daily MINIMUM prices using TSN (180-60-10 windows).}
  \label{fig:decomposition_plot_minimum}
\end{figure}

Figure~\ref{fig:decomposition_plot_minimum} shows the corresponding causal TSN decomposition applied to daily retail electricity \emph{minimum} prices, using comparable cutoff horizons of 180, 60, and 10 days. The semiannual component highlights persistent low-price regimes, whereas the quarterly and monthly components capture more abrupt cyclical movements and rapid adjustments that are characteristic of minimum-price series. The fluctuation component reveals sharp, short-lived price drops and spikes, which are more pronounced for minimum prices due to competitive pricing behavior and promotional offers.

Overall, the two decompositions illustrate that retail electricity prices exhibit distinct dynamics across temporal scales and price definitions. The resulting multi-scale components are interpretable in the sense that each corresponds to a clearly defined time horizon and contributes additively to observed prices, allowing variation to be directly attributed to long-run trends, intermediate cycles, or short-term fluctuations. These horizon-specific features are therefore well suited for downstream forecasting and causal analysis in retail electricity markets.

\subsection{Forecast accuracy and stability}

Table~\ref{tab:final-test-results_median} reports the forecast performance of alternative models for retail electricity daily \emph{median} prices. Overall, the proposed CG-TCN model achieves strong and consistent performance across all evaluation metrics, with low MAE (0.2685), MSE (0.1302), and RMSE (0.3609), as well as a small NRMSE (6.49\%) and a high coefficient of determination ($R^2=0.9648$). Among the benchmark models, Autoformer delivers slightly lower MAE and MAPE, but exhibits higher RMSE and lower overall stability. MTGNN also performs competitively, though with larger errors across all metrics. In contrast, conventional machine-learning models such as RF and XGB exhibit substantially higher forecast errors and limited explanatory power, while the classical AR model performs poorly with a negative $R^2$. Deep learning baselines including LSTM, TCN, and CTCN improve upon the machine-learning approaches but remain less accurate than CG-TCN. Overall, these results indicate that incorporating causal graph information into temporal convolutional networks yields robust gains in forecasting median retail electricity prices. Details of the hyperparameter optimization procedure and the selected configurations for CG-TCN and baseline deep-learning models are provided in~\ref{sec:appendix_parameter}.

\begin{table}[htbp]
\centering
\caption{Forecast results of retail electricity daily MEDIAN prices.}
\label{tab:final-test-results_median}
\small
\begin{tabular}{l
                S[table-format=1.4]
                S[table-format=1.4]
                S[table-format=1.4]
                S[table-format=3.4, table-space-text-post=\%]
                S[table-format=2.4, table-space-text-post=\%]
                S[table-format=1.4]}
\toprule
\multicolumn{1}{c}{\textbf{Model}} &  \multicolumn{1}{c}{\textbf{MAE}} & \multicolumn{1}{c}{\textbf{MSE}} & \multicolumn{1}{c}{\textbf{RMSE}} & \multicolumn{1}{c}{\textbf{MAPE}} & \multicolumn{1}{c}{\textbf{NRMSE}} & \multicolumn{1}{c}{$\mathbf{R^2}$} \\
\midrule
CG-TCN       &  0.2685 & \bfseries 0.1302 & \bfseries 0.3609 & 3.08\% &  6.49\% & \bfseries 0.9648 \\
\addlinespace
Autoformer         & \bfseries 0.2437 & 0.1710 & 0.4135 & \bfseries 2.53\% & \bfseries 6.21\% & 0.9627 \\
MTGNN              & 0.3196 & 0.2743 & 0.5238 & 3.33\% & 7.87\% & 0.9402 \\
\addlinespace
RF                 & 0.7761 & 1.3439 & 1.1593 & 117.32\% & 10.97\% & 0.3536 \\
XGB                & 0.7947 & 1.3842 & 1.1765 & 138.52\% & 11.13\% & 0.3343 \\
AR                 & 1.3265 & 2.9334 & 1.7127 & 170.01\% & 16.21\% & -0.4108 \\
LSTM               & 1.0135 & 1.8168 & 1.3479 & 183.31\% & 12.75\% & 0.1313 \\
TCN                & 0.8620 & 1.2352 & 1.1114 & 177.73\% & 10.52\% & 0.4094 \\
\addlinespace
CTCN               & 0.5935 & 0.7305 & 0.8547 & 250.51\% & 8.09\% & 0.6877 \\
\bottomrule
\end{tabular}
\end{table}

Table~\ref{tab:final-test-results_min} summarizes the corresponding forecast results for retail electricity daily \emph{minimum} prices. Because the minimum-price series contains zero values, the MAPE is not reported, as it becomes undefined or unstable in the presence of zeros. Focusing on scale-dependent metrics, CG-TCN again achieves the best overall performance, with the lowest MAE (0.2772), MSE (0.1534), RMSE (0.3917), and NRMSE (4.04\%), together with the highest $R^2$ (0.9693). Competing deep learning models, including Autoformer and MTGNN, show reasonable accuracy but consistently underperform CG-TCN, particularly in terms of RMSE and explained variance. Traditional machine-learning methods and the AR benchmark exhibit markedly higher errors and lower explanatory power. These results suggest that the benefits of the proposed causal-graph-enhanced architecture extend to the more volatile and discontinuous minimum-price series, where short-horizon dynamics and abrupt price movements are especially pronounced.

\begin{table}[htbp]
\centering
\caption{Forecast results of retail electricity daily MINIMUM prices.}
\label{tab:final-test-results_min}
\small
\begin{tabular}{l
                S[table-format=1.4]
                S[table-format=1.4]
                S[table-format=1.4]
                % S[table-format=3.4, table-space-text-post=\%]
                S[table-format=2.4, table-space-text-post=\%]
                S[table-format=1.4]}
\toprule
\multicolumn{1}{c}{\textbf{Model}} & \multicolumn{1}{c}{\textbf{MAE}} & \multicolumn{1}{c}{\textbf{MSE}} & \multicolumn{1}{c}{\textbf{RMSE}} & \multicolumn{1}{c}{\textbf{NRMSE}} & \multicolumn{1}{c}{$\mathbf{R^2}$} \\
\midrule
CG-TCN      & \bfseries 0.2772 & \bfseries 0.1534 & \bfseries 0.3917 & \bfseries 4.04\% & \bfseries 0.9693 \\
\addlinespace
Autoformer         & 0.3126 & 0.2696 & 0.5192 & 7.80\% & 0.9413 \\
MTGNN              & 0.3094 & 0.2891 & 0.5377 & 8.08\% & 0.9370 \\
\addlinespace
RF                 & 0.6849 & 1.8430 & 1.3576 & 5.11\% & 0.5329 \\
XGB                & 0.7214 & 2.2044 & 1.4847 & 5.59\% & 0.4413 \\
AR                 & 1.4962 & 4.7231 & 2.1733 & 8.18\% & -0.1970 \\
LSTM               & 0.9743 & 2.4621 & 1.5691 & 5.91\% & 0.3793 \\
TCN                & 0.85870 & 1.8145 & 1.3470 & 5.07\% & 0.5426 \\
\addlinespace
CTCN               & 0.6231 & 1.2681 & 1.1261 & 4.24\% & 0.6803 \\
\bottomrule
\end{tabular}
\end{table}

\begin{figure}[htbp]
  \centering
  \includegraphics[width=0.75\linewidth]{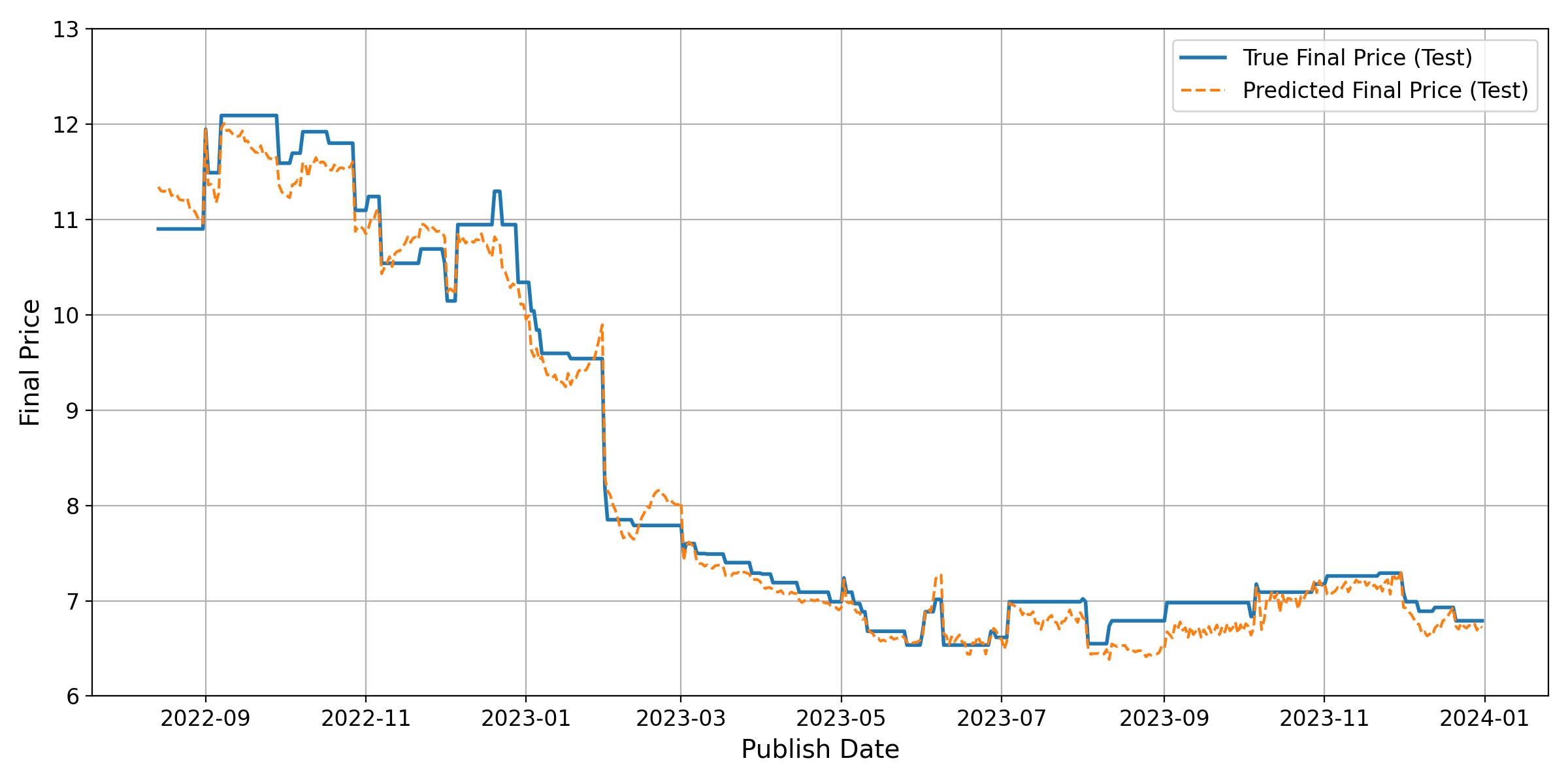}
  \caption{CG-TCN model predictions of daily median retail electricity prices during the test period.}
  \label{fig:test_final}
\end{figure}

Figure~\ref{fig:test_final} illustrates the out-of-sample test performance of the best-performing CG-TCN model by comparing the predicted daily retail electricity median prices with the observed values. The predicted trajectory closely follows the overall downward price movement during the test period and remains well aligned with both gradual regime shifts and pronounced price drops. While small discrepancies arise during periods of heightened short-term volatility, the model accurately reproduces the dominant price dynamics over the entire horizon.

\begin{figure}[htbp]
  \centering
  \includegraphics[width=0.7\linewidth]{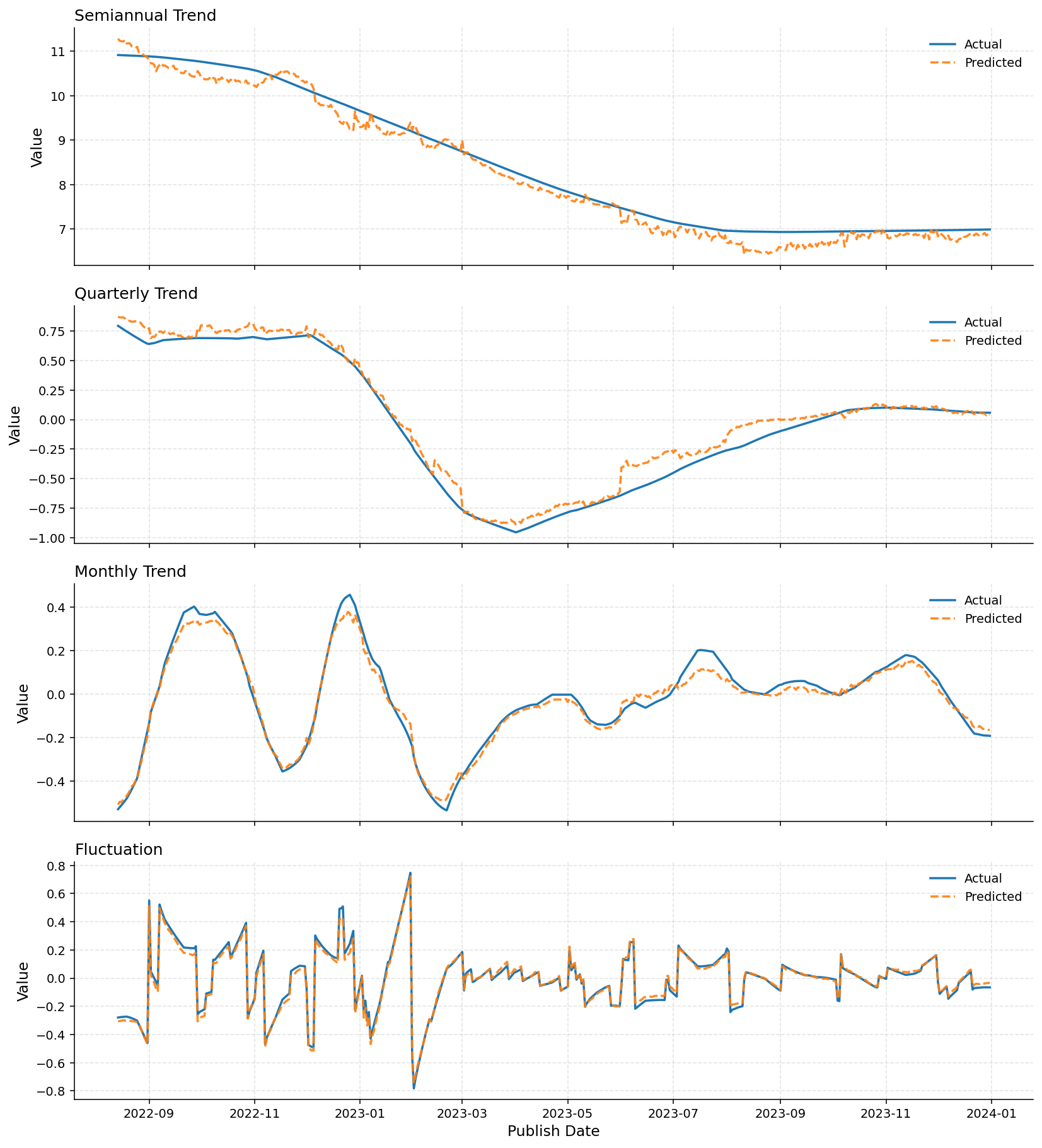}
  \caption{CG-TCN predictions of multi-scale decomposition components of daily median retail electricity prices during the test period.}
  \label{fig:test_component}
\end{figure}

To further examine the sources of predictive accuracy, Figure~\ref{fig:test_component} presents the corresponding predictions for the decomposed multi-scale components. The semiannual component demonstrates that the model effectively captures the long-run downward trend in prices, while the quarterly component closely tracks medium-term adjustments associated with evolving market conditions. The monthly component shows strong alignment between predicted and observed short-horizon oscillations, indicating the model’s ability to adapt to shorter-term cyclical behavior. Finally, the fluctuation component reveals that the model is able to reproduce high-frequency variability, although extreme spikes are slightly attenuated, reflecting a modest smoothing effect. Taken together, these results indicate that the proposed CG-TCN framework not only achieves accurate aggregate price forecasts but also reliably represents price dynamics across multiple temporal scales.

\subsection{Multi-step-ahead forecasting results}

Table~\ref{tab:multistep_results} reports the multi-step forecasting performance for daily median prices across horizons $H=\{5,10,15\}$. Multi-step forecasting is particularly relevant in retail electricity markets, where pricing, procurement, and contract design decisions often rely on price expectations several days to weeks ahead rather than on one-step-ahead forecasts. Evaluating performance at longer horizons therefore provides a more stringent and operationally meaningful assessment of model robustness. For each horizon, models are evaluated using direct $H$-step-ahead forecasts, with predictions aligned to observed prices at $t+H$ and metrics computed on a held-out test set, thereby avoiding recursive error accumulation.

% Across all horizons, CG-TCN consistently achieves the lowest MAE, MSE, and RMSE, as well as the highest $R^2$, indicating superior accuracy and explanatory power relative to the benchmark models. As expected, forecast errors increase with horizon length for all approaches; however, the degradation is notably less pronounced for CG-TCN, suggesting greater robustness to error accumulation in multi-step forecasting.

Across all horizons, CG-TCN consistently outperforms the benchmark models. For example, at the 5-day horizon, CG-TCN achieves an MAE of 0.404 and an $R^2$ of 0.935, compared with an MAE of 0.546 and $R^2$ of 0.840 for Autoformer. This performance gap persists and widens at longer horizons: at $H=10$, CG-TCN maintains a low RMSE of 0.499 and $R^2=0.934$, while competing models exhibit substantially higher errors and lower explanatory power. At the 15-day horizon, although forecast errors increase for all approaches, CG-TCN continues to deliver the lowest MAE and RMSE and the highest $R^2$, indicating greater robustness to error accumulation in multi-step forecasting.

\begin{table}[htbp]
\centering
\caption{Multi-step forecasting for daily median price across horizons $H\in\{5,10,15\}$.}
\label{tab:multistep_results}
\small
\begin{tabular}{l l r r r r r r r}
\toprule
Horizon & Model  & MAE & MSE & RMSE & MAPE & NRMSE & $R^2$ \\ \midrule
\multirow{3}{*}{5}  & CG-TCN   & \textbf{0.4044} & \textbf{0.2428} & \textbf{0.4927} & \textbf{4.73\%} & \textbf{8.87\%}  & \textbf{0.9347} \\
                    & Autoformer  & 0.5460 & 0.7358 & 0.8578 & 5.73\% & 12.89\% & 0.8398 \\
                    & MTGNN      & 0.5739 & 0.7968 & 0.8927 & 6.16\% & 13.41\% & 0.8263 \\

\addlinespace
\multirow{3}{*}{10} & CG-TCN    & \textbf{0.3548} & \textbf{0.2485} & \textbf{0.4985} & \textbf{3.82\%} & \textbf{8.97\%} & \textbf{0.9335} \\
                    & Autoformer & 0.6112 & 0.8418 & 0.9175 & 6.49\% & 13.79\% & 0.8170 \\
                    & MTGNN      & 0.7297 & 1.1434 & 1.0693 & 7.69\% & 16.07\% & 0.7507 \\

\addlinespace
\multirow{3}{*}{15} & CG-TCN     & \textbf{0.5419} & \textbf{0.7337} & \textbf{0.8565} & \textbf{5.43\%} & \textbf{15.41\%} & \textbf{0.8052} \\
                    & Autoformer & 0.6558 & 0.9428 & 0.9710 & 6.88\% & 14.59\% & 0.7953 \\
                    & MTGNN      & 0.7574 & 1.3702 & 1.1706 & 8.10\% & 17.59\% & 0.7013 \\

\addlinespace
\bottomrule
\end{tabular}
\end{table}

Autoformer and MTGNN are selected as baseline models because they represent state-of-the-art deep learning architectures for long-sequence and multivariate time-series forecasting and demonstrate strong one-step-ahead performance in prior studies and in our empirical setting. Limiting the comparison to these competitive baselines allows for a focused evaluation of multi-step performance without conflating results with models that already underperform in the short-horizon setting. In contrast to these benchmarks, CG-TCN maintains more stable accuracy as the prediction horizon increases, highlighting the advantages of its causal and multi-scale design for capturing persistent temporal structure in retail electricity prices.

\subsection{Explainability result}

\subsubsection{Causal DAG}
\label{sec:causal_DAG}
% Beyond predictive accuracy, interpretability is a central requirement for causal analysis. And, this is particularly policy relevant in retail electricity markets, where understanding the mechanisms driving price formation is essential for economic interpretation and policy relevance. And this is even more relevant given the current national debate in the US regarding energy affordability.

%In this context, causal graphs provide a transparent representation of how price components at different temporal scales relate to contract attributes, wholesale market conditions, and exogenous factors. Such structured representations allow us to distinguish persistent drivers from short-term fluctuations, assess whether learned dependencies align with economic theory, and avoid spurious feedback loops that can arise in purely data-driven models.

Beyond predictive accuracy, interpretability is a central requirement for critical applications, particularly in retail electricity markets where policy design, consumer protection, and regulatory oversight depend on understanding the mechanisms driving price formation. As discussed in Sections~1 and~2, much of the existing forecasting literature relies on black-box machine-learning models that deliver high predictive accuracy but provide limited insight into why prices evolve as they do, limiting their usefulness for economic interpretation and policy analysis.

In this context, causal graphs serve as a critical methodological bridge between predictive performance and economic explainability. By explicitly representing directed dependencies among decomposed price components, contract attributes, wholesale market conditions, and exogenous factors, causal DAGs allow us to assess whether learned relationships align with economic theory, institutional knowledge, and the unidirectional nature of price formation. This addresses a key gap in prior work on retail electricity price forecasting, which has rarely integrated formal causal discovery with interpretable multi-scale price decomposition.

\begin{figure}[htbp]
  \centering
  \includegraphics[width=0.8\linewidth]{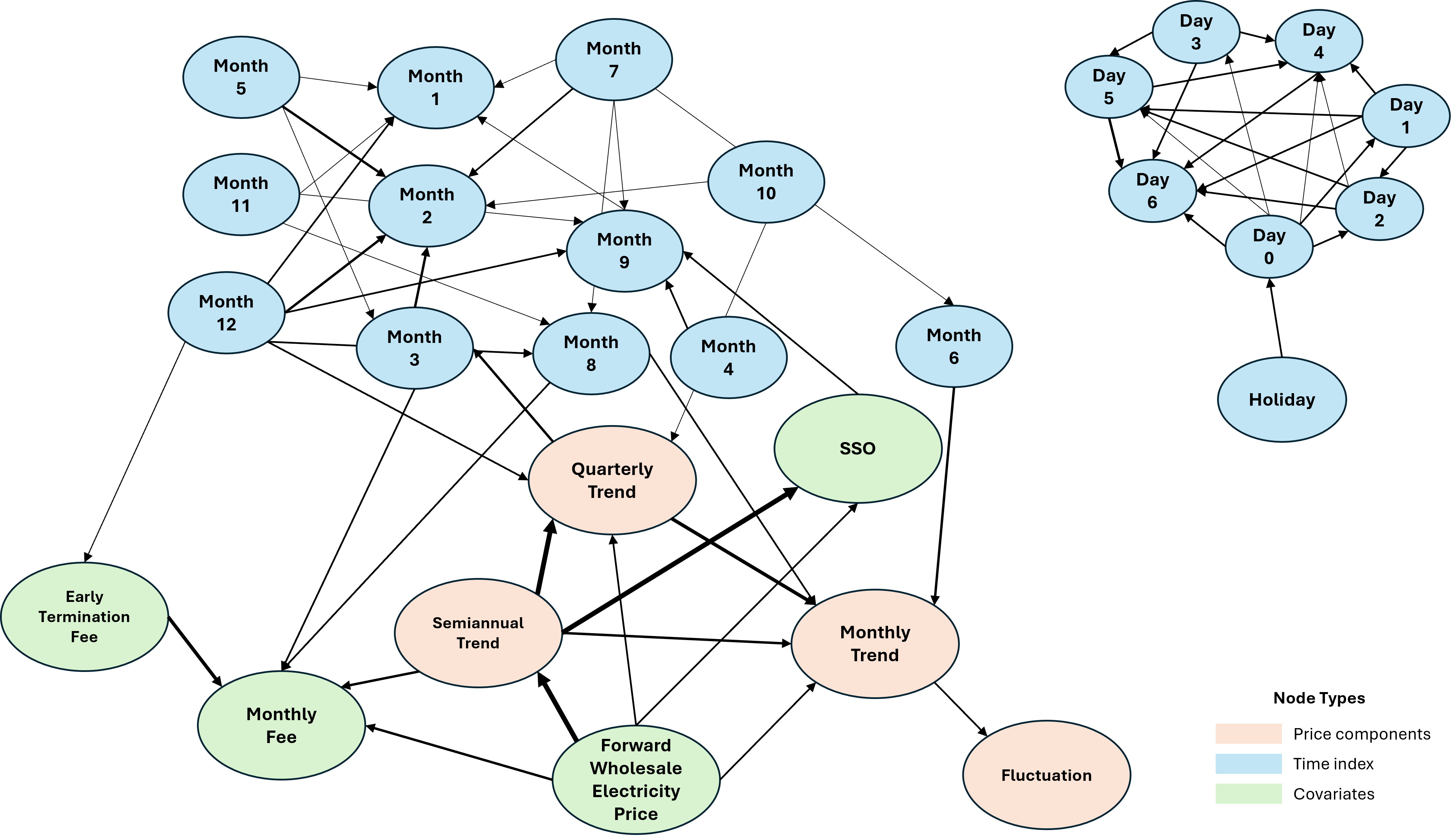}
  \caption{Estimated causal graph without structural constraints.}
  \label{fig:causal_graph_noconstraint}
\end{figure}

\begin{figure}[htbp]
  \centering
  \includegraphics[width=0.8\linewidth]{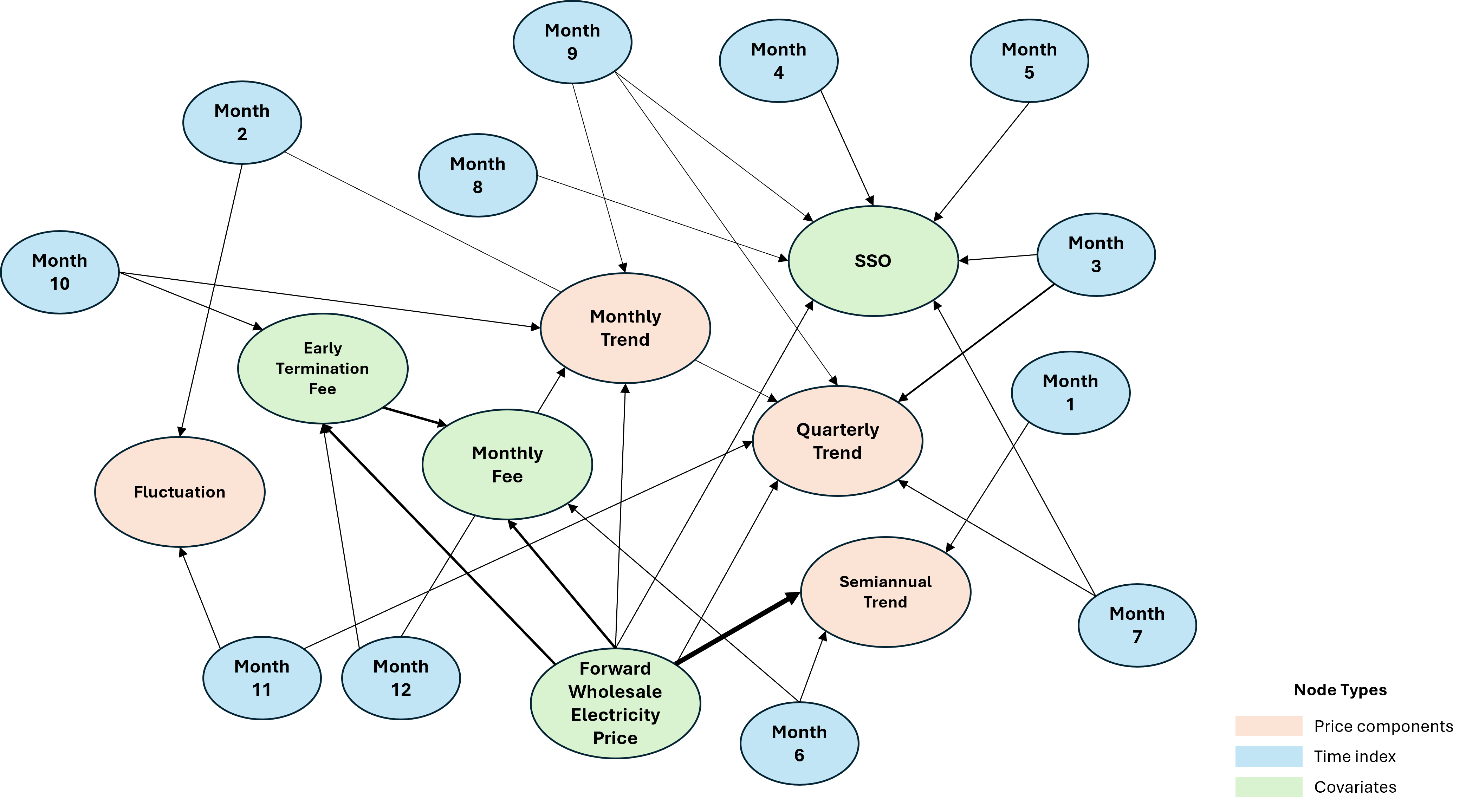}
  \caption{Estimated causal graph with forecasting-target and exogenous-input constraints.}
  \label{fig:causal_graph_constraint}
\end{figure}

Figures~\ref{fig:causal_graph_noconstraint} and~\ref{fig:causal_graph_constraint} compare the estimated causal structures among price components, contract attributes, covariates, and time features under unconstrained and constrained learning settings. In the unconstrained graph (Figure~\ref{fig:causal_graph_noconstraint}), dense interconnections emerge, particularly among monthly and daily time features, 
%resulting in a highly entangled network in which forecasting targets (i.e., semiannual, quarterly, monthly, and fluctuation components) appear both as parents and children. Such bidirectional dependencies complicate interpretation and are difficult to reconcile with the unidirectional nature of causal price formation.
resulting in a highly entangled network. In this setting, price components at different temporal scales are simultaneously inferred to influence one another and to be influenced by other variables, creating feedback-like relationships with no clear causal ordering. Such structures complicate interpretation and are difficult to reconcile with the unidirectional and temporally ordered nature of causal price formation in retail electricity markets.

By contrast, the constrained graph (Figure~\ref{fig:causal_graph_constraint}) enforces two economically motivated structural realities: (1) forecasting targets cannot act as parent nodes, and (2) time features and wholesale forward prices cannot appear as causal child nodes. These constraints yield a sparser and more interpretable network, in which causal directions align more closely with theoretical expectations—price components depend on external covariates and market inputs, while exogenous features remain independent drivers. This comparison highlights the importance of incorporating domain knowledge into causal discovery to obtain graphs that are both statistically coherent and economically meaningful in retail electricity markets.

Importantly, this comparison is not merely descriptive. It demonstrates how unconstrained, purely data-driven causal discovery can produce statistically admissible but economically implausible structures, reinforcing concerns raised in the causal interpretability literature regarding spurious feedback loops and ambiguous causal ordering in high-dimensional time-series settings \cite{pfister2018kernel, scholkopf2021toward}. By introducing economically motivated structural constraints, our approach embeds domain knowledge directly into the causal discovery process, yielding graphs that are not only statistically coherent but also economically interpretable. In this sense, the constrained DAG constitutes a substantive methodological contribution: it operationalizes explainability in causal forecasting by aligning learned structures with institutional and theoretical expectations rather than relying solely on post hoc interpretation.

The causal structures are estimated using the NOTEARS framework, which provides a differentiable approach to learning directed acyclic graphs (DAGs) by enforcing a smooth acyclicity constraint. In the linear NOTEARS formulation \cite{zheng2018dags}, each variable is modeled as a linear function of its potential parents, and the optimization jointly estimates regression coefficients while constraining the resulting adjacency matrix to be acyclic. This approach yields sparse and transparent causal graphs that are straightforward to interpret. The nonlinear extension of NOTEARS replaces linear regressions with neural networks at each node, allowing for flexible nonlinear dependencies while maintaining the same acyclicity constraint \cite{zheng2020learning}. As a result, the two variants offer a natural trade-off between interpretability and expressive power.

\begin{figure}[htbp]
  \centering
  \includegraphics[width=0.8\linewidth]{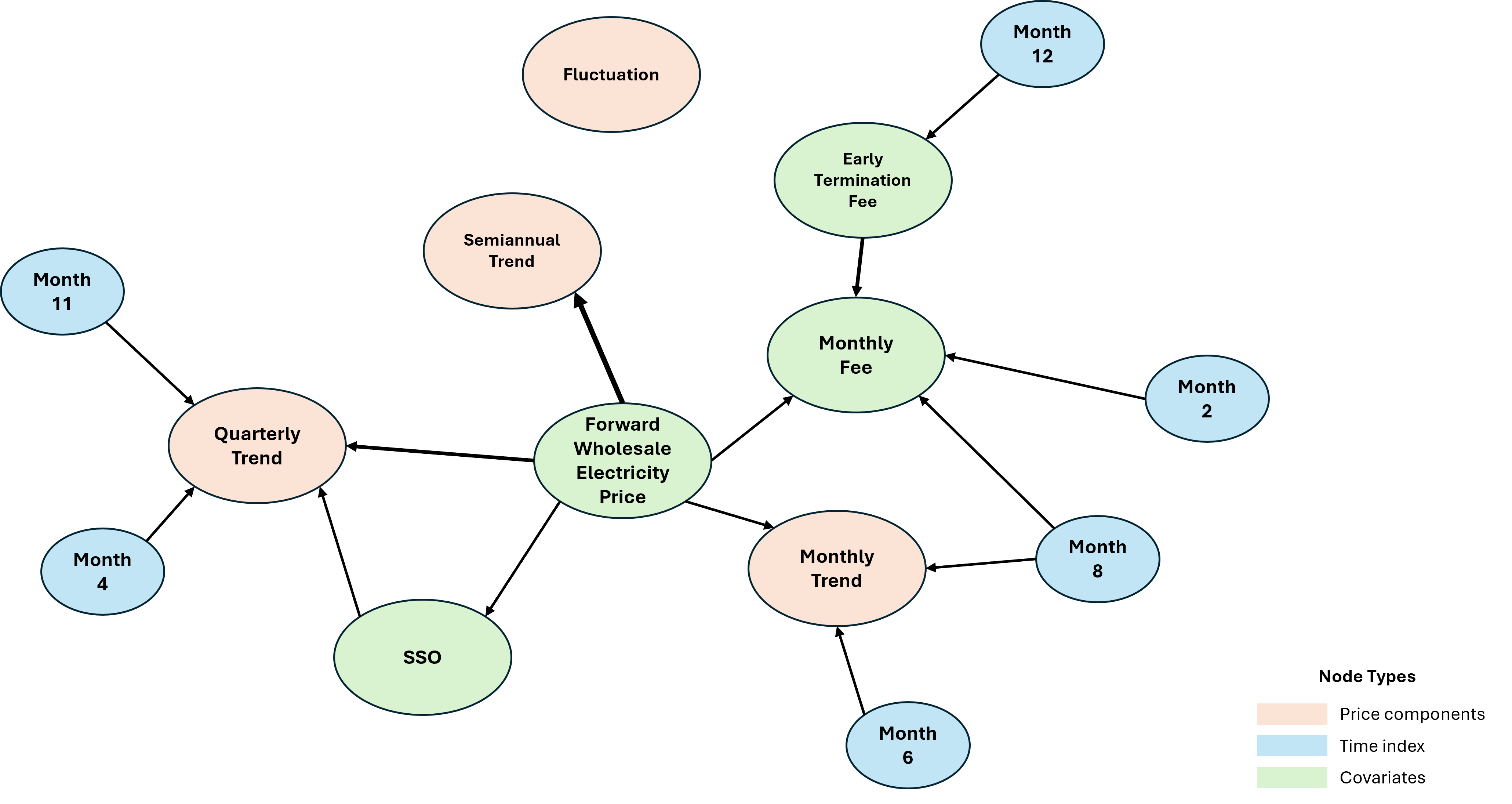}
  \caption{Estimated causal graph from the nonlinear NOTEARS algorithm.}
  \label{fig:causal_graph_nonlinear}
\end{figure}

Figures~\ref{fig:causal_graph_constraint} and~\ref{fig:causal_graph_nonlinear} illustrate the causal graphs learned by the linear and nonlinear NOTEARS algorithms under the constrained setting. The linear NOTEARS graph is relatively dense, with multiple connections among price components (red nodes), contract attributes and covariates (green nodes), and time features (blue nodes). This reflects the tendency of linear models to capture a broad set of associations, including weaker but potentially informative relationships. In contrast, the nonlinear NOTEARS graph is more selective and parsimonious, highlighting a smaller number of strong causal links—most notably between wholesale forward prices, contract fees, and key price trend components. This sparsity suggests that the nonlinear formulation is better able to focus on economically meaningful dependencies while filtering out weaker or spurious correlations.

Table~\ref{tab:notears_linear_vs_nonlinear} compares the forecasting performance for daily median prices of the CG-TCN model when using causal graphs derived from linear and nonlinear NOTEARS. Overall, the two variants achieve very similar predictive accuracy. The nonlinear NOTEARS graph yields marginal improvements in $R^2$ and RMSE/MSE on the test set, whereas the linear NOTEARS graph achieves slightly lower MAE and MAPE. This pattern reflects the balance between model flexibility and data availability. While the nonlinear NOTEARS formulation introduces additional parameters that can capture subtle nonlinear dependencies, the available sample size (approximately 3,600 daily observations) limits the extent to which this additional capacity translates into systematic gains. By contrast, the simpler linear NOTEARS approach provides a more stable, lower-variance representation under the current data regime. These results suggest that nonlinear NOTEARS may offer clearer advantages when larger datasets, stronger regularization, or more restrictive domain-informed constraints are available. In this study, we adopt the \emph{linear} NOTEARS formulation for causal graph discovery, using the nonlinear variant only for comparative and robustness analysis. A comparison between the saliency-based causal interpretation of the baseline CTCN and the explicit causal structure learned by CG–TCN is provided in~\ref{sec:appendix_ctcn_comparison}.

\begin{table}[htbp]
\centering
\caption{Performance of CG-TCN using causal graphs learned from linear and nonlinear NOTEARS. The two variants exhibit comparable predictive accuracy.}
\label{tab:notears_linear_vs_nonlinear}
\small
\begin{tabular}{l r r r r r r}
\toprule
Model & MAE & MSE & RMSE & MAPE & NRMSE & $R^2$ \\
\midrule
CG-TCN (Linear NOTEARS)
&  \bfseries 0.2685 &  0.1302 & 0.3609 & \bfseries 3.08\% &  6.49\% &\bfseries 0.9648 \\
CG-TCN (Nonlinear NOTEARS)
& 0.2841 & \textbf{0.1291} & \textbf{0.3212} & 3.61\% & \textbf{5.06\%} & 0.9556 \\
\bottomrule
\end{tabular}
\end{table}

Although the causal graphs estimated using the NOTEARS framework uncover meaningful dependencies among contract attributes, wholesale price signals, and decomposed price components, some target components lack clearly identified explanatory parents in these static DAGs. This limitation stems from the sparsity-inducing nature of NOTEARS, which is designed to retain only statistically dominant direct edges while pruning weaker, indirect, or distributed relationships \cite{ng2020role, liu2024cl}. While such sparsity enhances interpretability, it may omit pathways that are collectively important for prediction. To address this limitation and improve forecasting performance, we embed the discovered causal structure into a Graph Neural Network (GNN). The GNN allows each node to aggregate information not only from its direct parents but also from higher-order neighbors, thereby capturing distributed, nonlinear, and cross-component interactions that static causal graphs alone cannot represent. This causal-graph-informed embedding preserves the interpretable backbone provided by NOTEARS while extending it into a flexible representation optimized for temporal forecasting, effectively bridging causal discovery and predictive modeling.

\begin{figure}[htbp]
  \centering
  \includegraphics[width=0.9\linewidth]{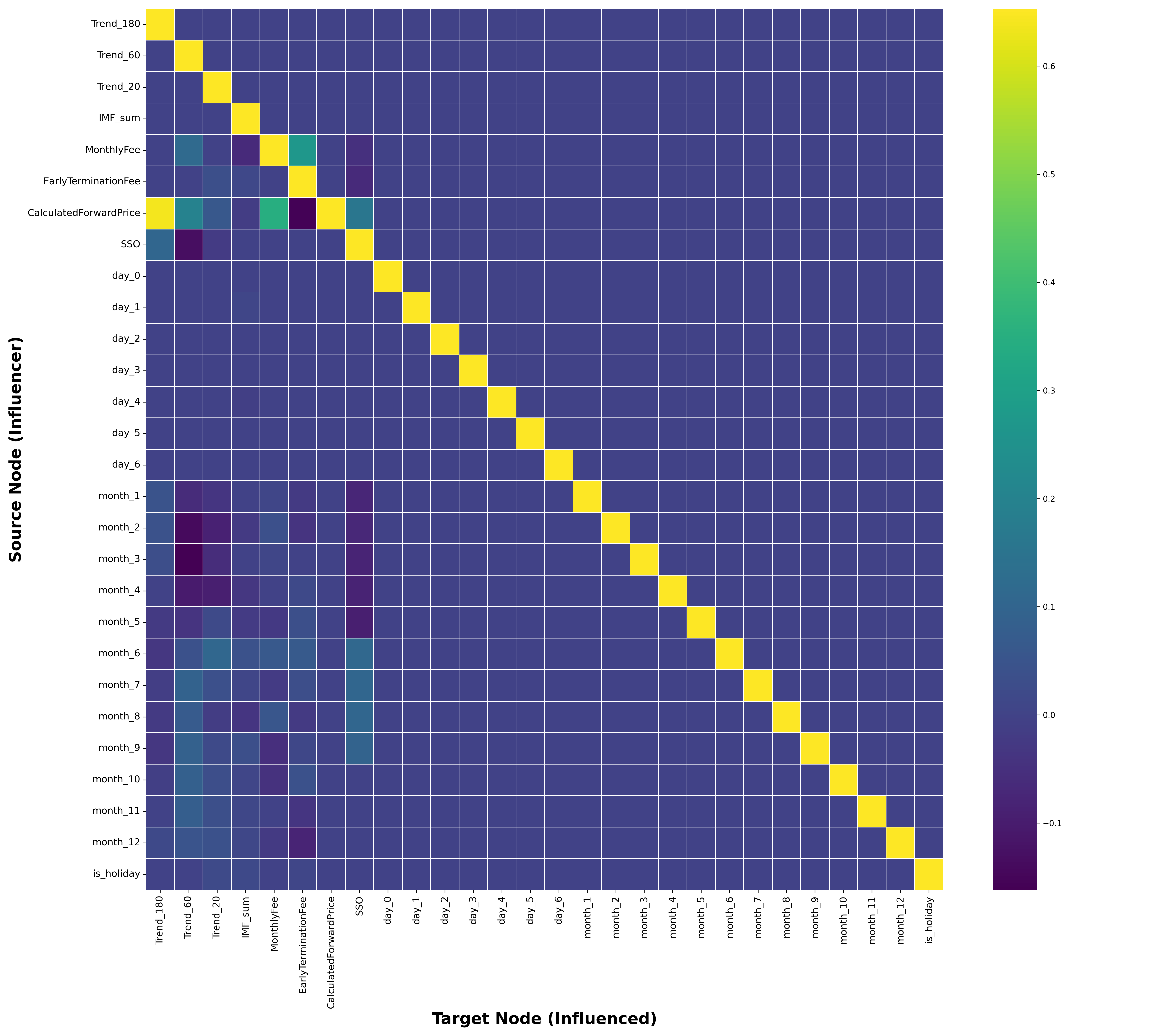}
  \caption{GNN-learned effective adjacency matrix representing causal influences among price components, covariates, and time features.}
  \label{fig:effective_adjacency}
\end{figure}

Figure~\ref{fig:effective_adjacency} presents the effective adjacency matrix learned by the GNN-based causal forecasting model. The matrix summarizes the learned directed influence strengths among all variables, with color intensity indicating the magnitude of each influence; diagonal entries are fixed at unity to reflect self-dependence. Several economically intuitive patterns emerge. The calculated wholesale forward price exhibits a strong directed influence on the long-horizon trend component (Trend\_180), indicating that expectations embedded in forward markets play a central role in shaping persistent retail price movements. In contrast, the monthly contract fee primarily influences the high-frequency fluctuation component (IMF\_sum), suggesting that supplier-specific pricing adjustments mainly affect short-term variability rather than long-run trends. The SSO price also appears as an important driver. Prior research has indicated that the SSO plays a key role as a price heuristic for the open market operations of retail suppliers, or CRES \cite{dormady2025efficiency}. This plays a role in linking the default price offered by the distribution utility to both trend and fluctuation components. By comparison, daily calendar indicators and holiday effects exhibit relatively weak influence, consistent with their limited role in determining retail electricity prices. 
%Overall, the learned adjacency structure reveals a hierarchical organization of causal influences, in which attach structural market signals govern long-term price evolution, while contract-level attributes modulate short-term dynamics. These results further demonstrate how the GNN embedding translates economically meaningful causal relationships into improved forecasting performance.
Overall, the learned adjacency structure reveals a hierarchical organization of causal influences, in which underlying structural market signals, such as wholesale forward prices and the SSO, govern long-term price evolution, while contract-level attributes primarily modulate short-term dynamics. These results further demonstrate how the GNN embedding translates economically meaningful causal relationships into improved forecasting performance.

This causal DAG analysis provides a structured and interpretable view of the mechanisms underlying retail electricity price formation. By explicitly modeling directed dependencies among decomposed price components, contract attributes, wholesale market signals, and time features, the DAG framework allows us to assess whether learned relationships align with economic intuition and institutional knowledge. Rather than treating price dynamics as a black box, the causal graphs clarify how long-term price trends are driven by persistent market fundamentals, while short-term fluctuations are shaped by contract-level pricing strategies and transient market conditions. Importantly, the use of causal discovery complements the forecasting task by constraining information flow to economically plausible pathways, thereby enhancing both interpretability and robustness. These insights motivate the subsequent integration of the learned causal structure into the forecasting model, where causal relationships serve not only as explanatory tools but also as inductive biases that improve predictive performance.

\subsubsection{Temporal causality}

While the causal DAG analysis in Section~\ref{sec:causal_DAG} identifies \emph{which} variables are structurally connected in retail electricity price formation, temporal causality addresses a complementary question: \emph{when} different inputs are most informative for forecasting price dynamics. Understanding the timing of explanatory influence is particularly important in retail electricity markets, where contract attributes, wholesale price signals, and exogenous factors operate over distinct temporal horizons. This section therefore examines how the proposed CG-TCN model allocates explanatory importance across historical time lags, revealing the temporal scales at which different variables exert their predictive influence.

\begin{figure}[htbp]
\centering
\includegraphics[width=0.9\linewidth]{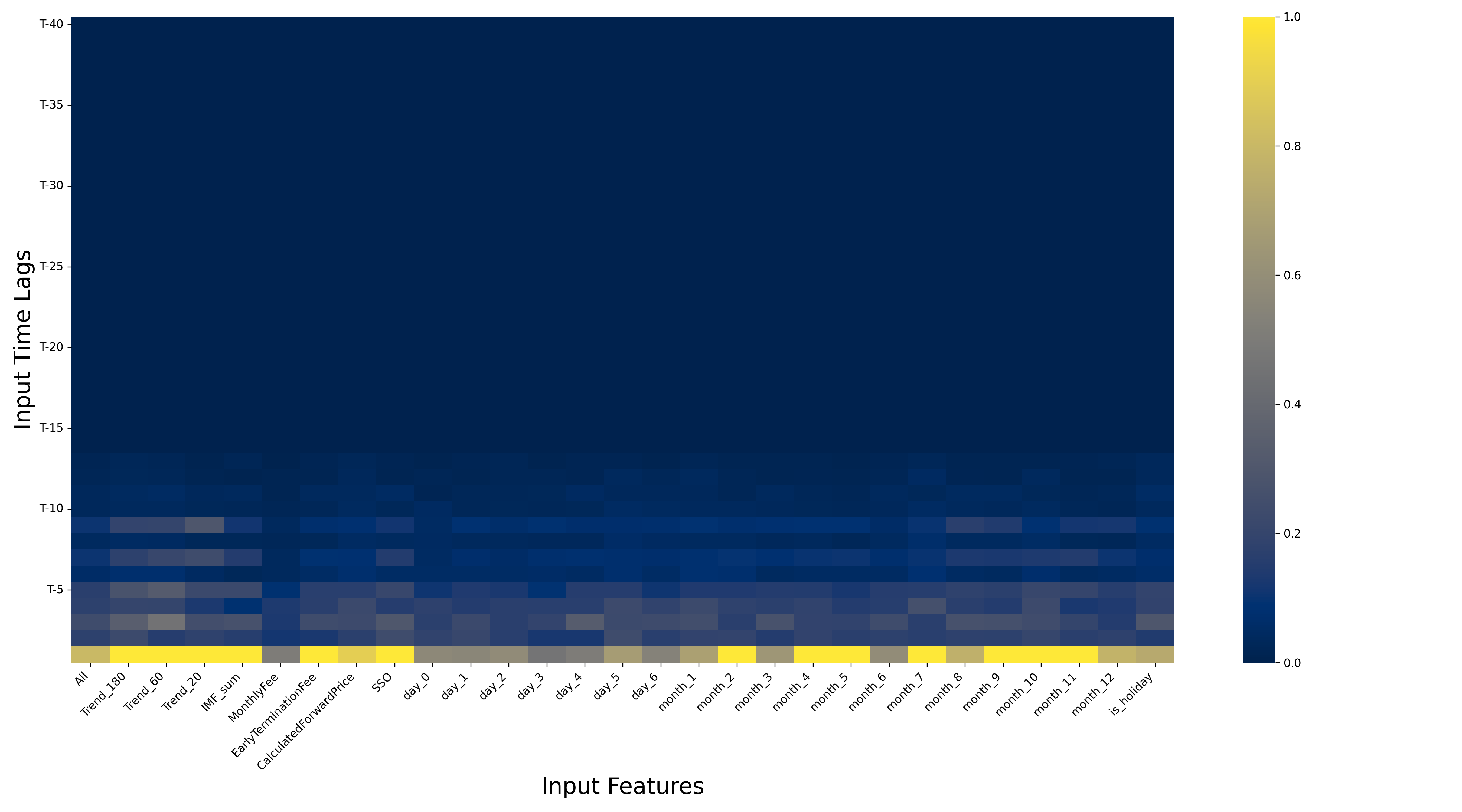}
\caption{Normalized temporal causality in forecasting the final retail electricity price. Each saliency value is normalized by the maximum magnitude across all time lags and features for comparability. The first column (“All”) indicates the average saliency across all input features at each time lag.}
\label{fig:timecausal_combined}
\end{figure}

Figures~\ref{fig:timecausal_combined} and~\ref{fig:timecausal_trends_panel} present normalized temporal saliency maps that visualize how the model allocates importance across historical time lags and input features when forming forecasts of the final retail electricity price and its decomposed components.
%Figures~\ref{fig:timecausal_combined} and~\ref{fig:timecausal_trends_panel} present normalized temporal saliency maps for forecasting the final retail electricity price and its decomposed components, including the semiannual, quarterly, and monthly trends, as well as the short-term fluctuation. 
Each heatmap visualizes how the model allocates importance across historical time lags and input features when forming predictions. The vertical axis corresponds to input time lags, ranging from recent observations (T--1) to longer-range dependencies (up to T--40), while the horizontal axis represents input features, including decomposed price components, contract attributes, and exogenous covariates. The first column (“All”) summarizes the average saliency across all input features at each time lag, providing an overall measure of temporal importance independent of feature type.

% \begin{figure}[htbp]
%   \centering

%   % Row 1
%   \begin{subfigure}{0.49\textwidth}
%     \centering
%     \includegraphics[width=\linewidth]{Figures/Fold_0_global_saliency_target_Test_Semiannual_Trend.png}
%     \caption{Semiannual trend}
%     \label{fig:timecausal_semiannual}
%   \end{subfigure}
%   \hfill
%   \begin{subfigure}{0.49\textwidth}
%     \centering
%     \includegraphics[width=\linewidth]{Figures/Fold_0_global_saliency_target_Test_Quarterly_Trend.png}
%     \caption{Quarterly trend}
%     \label{fig:timecausal_quarterly}
%   \end{subfigure}

%   \vspace{0.6em}

%   % Row 2
%   \begin{subfigure}{0.49\textwidth}
%     \centering
%     \includegraphics[width=\linewidth]{Figures/Fold_0_global_saliency_target_Test_Monthly_Trend.png}
%     \caption{Monthly trend}
%     \label{fig:timecausal_monthly}
%   \end{subfigure}
%   \hfill
%   \begin{subfigure}{0.49\textwidth}
%     \centering
%     \includegraphics[width=\linewidth]{Figures/Fold_0_global_saliency_target_Test_Fluctuation.png}
%     \caption{Fluctuation}
%     \label{fig:timecausal_fluctuation}
%   \end{subfigure}

%   \caption{Normalized temporal causality for decomposed components of retail electricity price. Each map is normalized to \([0,1]\) by its global maximum saliency. The first column (“All”) shows the average saliency across all input features for each time lag.}
%   \label{fig:timecausal_trends_panel}
% \end{figure}

% --- Page 1: (a)(b)
\begin{figure}[p]
  \centering

  \begin{subfigure}{\textwidth}
    \centering
    \includegraphics[width=0.88\linewidth]{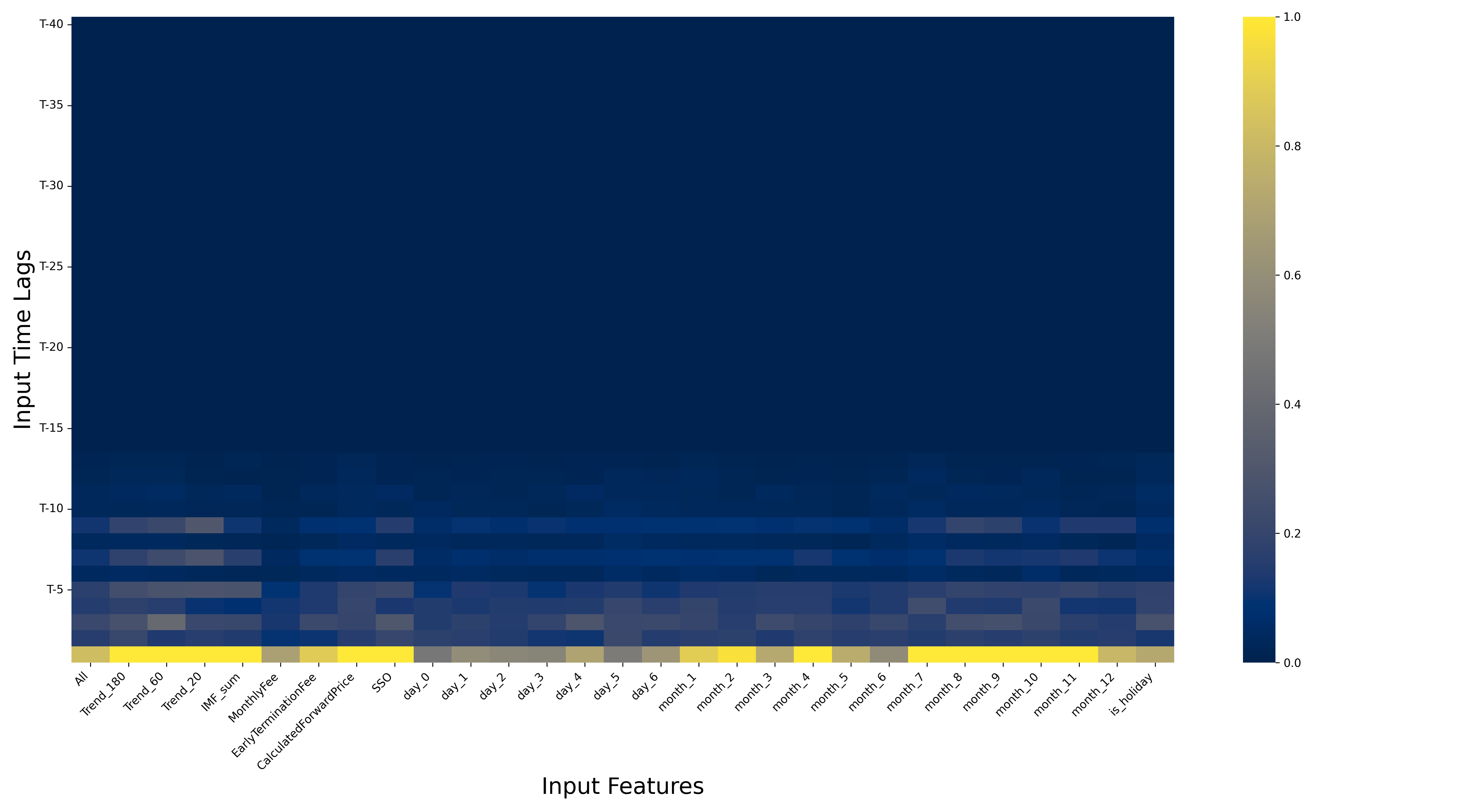}
    \caption{Semiannual trend}
    \label{fig:timecausal_semiannual}
  \end{subfigure}

  \vspace{0.8em}

  \begin{subfigure}{\textwidth}
    \centering
    \includegraphics[width=0.88\linewidth]{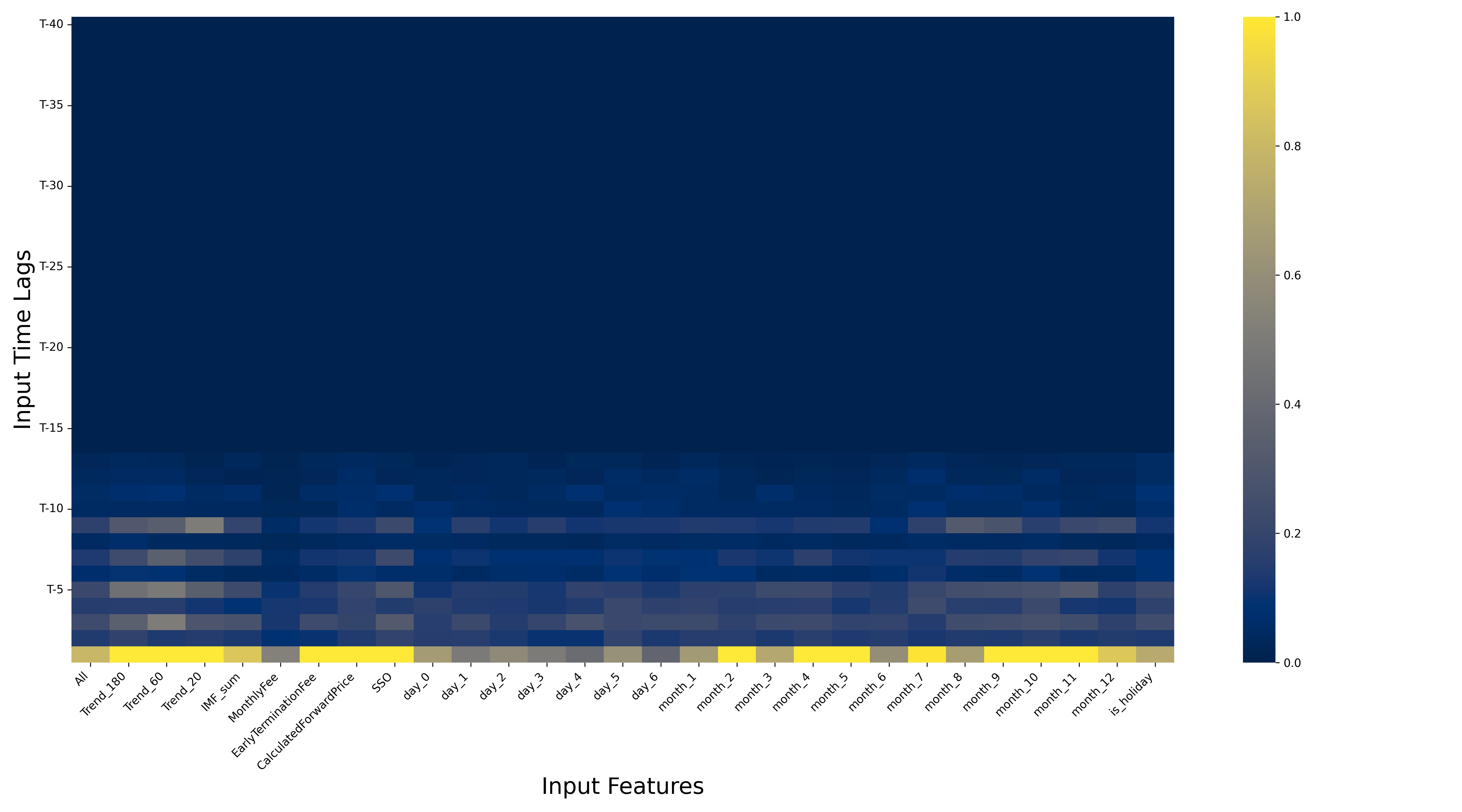}
    \caption{Quarterly trend}
    \label{fig:timecausal_quarterly}
  \end{subfigure}

  \caption{Normalized temporal causality for decomposed components of retail electricity price. Each map is normalized to \([0,1]\) by its global maximum saliency. The first column (“All”) shows the average saliency across all input features for each time lag.}
  \label{fig:timecausal_trends_panel}
\end{figure}

% --- Page 2: (c)(d_bt) continued
\begin{figure}[p]
  \ContinuedFloat
  \centering

  \begin{subfigure}{\textwidth}
    \centering
    \includegraphics[width=0.88\linewidth]{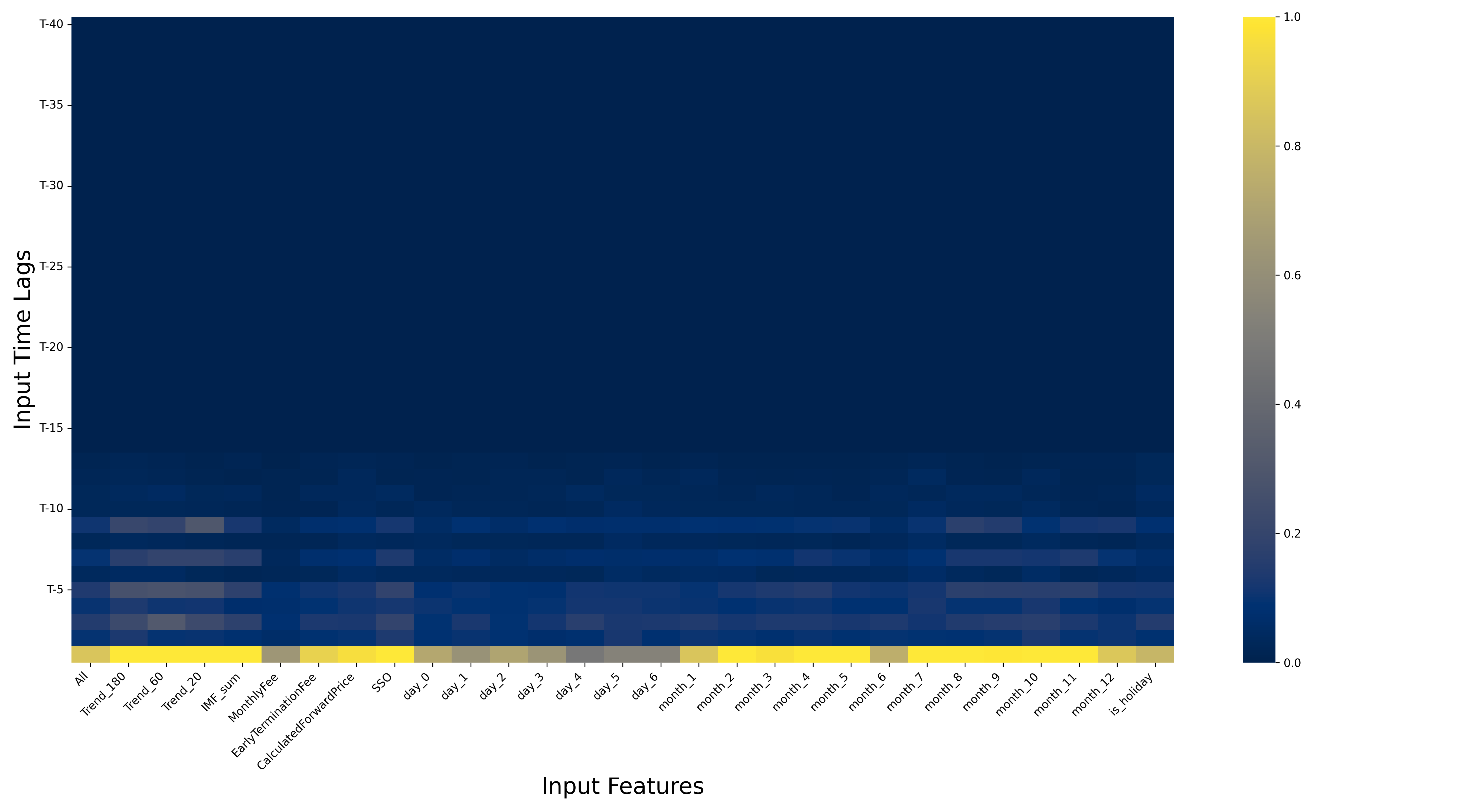}
    \caption{Monthly trend}
    \label{fig:timecausal_monthly}
  \end{subfigure}

  \vspace{0.8em}

  \begin{subfigure}{\textwidth}
    \centering
    \includegraphics[width=0.88\linewidth]{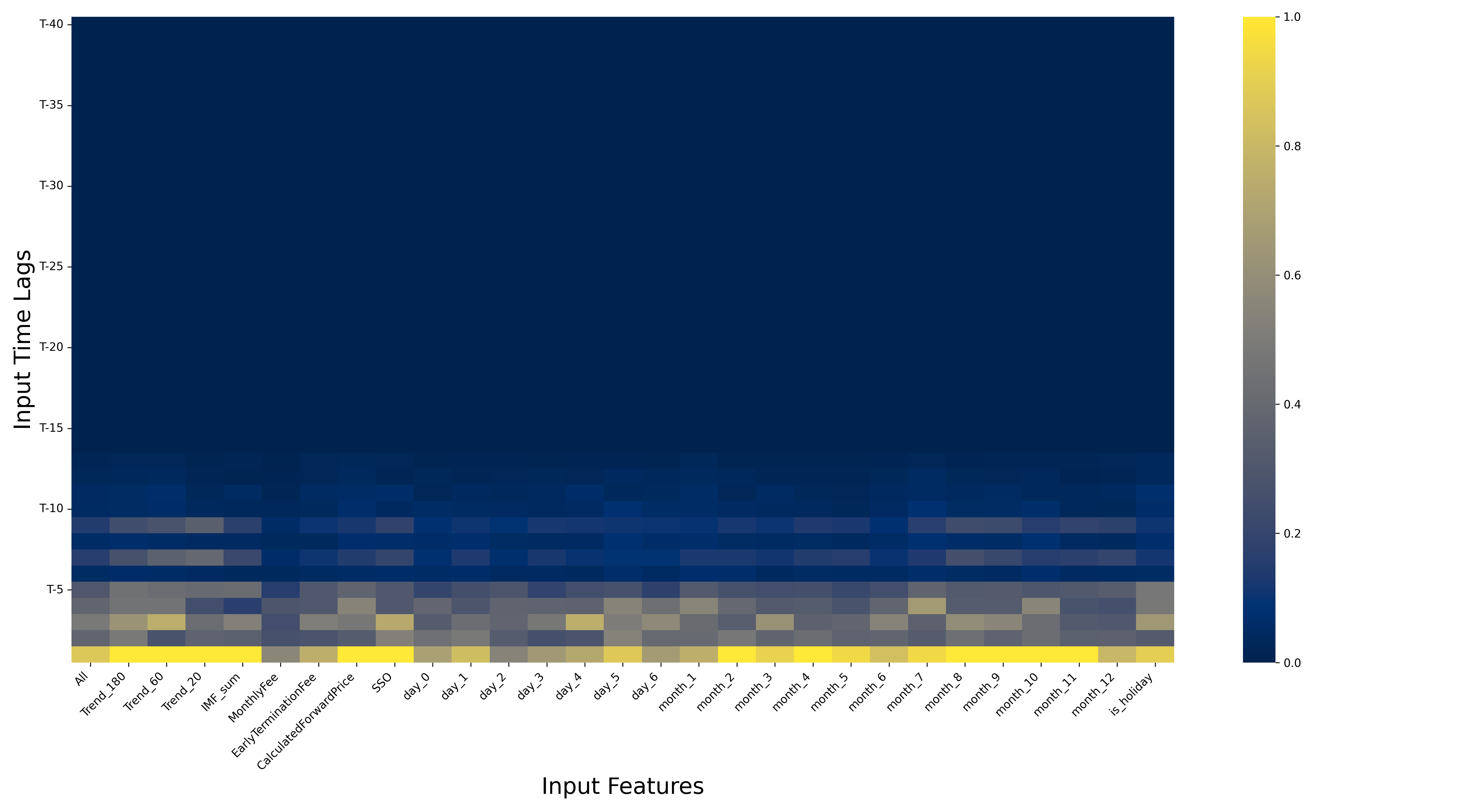}
    \caption{Fluctuation}
    \label{fig:timecausal_fluctuation}
  \end{subfigure}

  \caption{(continued)}
\end{figure}

To facilitate comparison across targets and features, all saliency values are normalized by the global maximum magnitude within each map, rescaling entries to the interval [0,1]. This normalization removes differences in absolute gradient scale and emphasizes relative temporal and feature-specific contributions within each forecasting task.

Across all targets, a consistent temporal pattern emerges: predictive influence is strongly concentrated in the most recent observations, particularly within lags T--1 to T--3, indicating that short-term dynamics play a dominant role in retail electricity price forecasting. Beyond these immediate lags, saliency rapidly attenuates, though select features retain moderate influence at longer horizons, reflecting persistent structural effects.

The distribution of saliency across features varies systematically by component. For the semiannual trend, pronounced and persistent saliency is observed for the calculated wholesale forward price and the long-horizon price trend (Trend\_180), consistent with their role as stabilizing long-term market signals. In contrast, the fluctuation component exhibits strong sensitivity to the monthly fee and the IMF\_sum index, suggesting that supplier contract terms and broader market conditions primarily drive short-term price variability. The quarterly and monthly trends display more diffuse saliency patterns, reflecting mixed dependence on recent contract attributes, seasonal indicators, and intermediate-term price dynamics.

When interpreted alongside the NOTEARS-derived causal graphs and the GNN-learned adjacency embeddings, these temporal saliency maps provide a dynamic complement to static causal structure learning \cite{pan2020series}. The causal DAGs identify which variables are structurally connected, while the GNN embedding refines these relationships into weighted directional links that govern information propagation. Temporal saliency further extends this framework by revealing \emph{when} these causal relationships exert the greatest influence on forecasts. Together, the static causal structure, learned graph embedding, and temporal causality analysis demonstrate that retail electricity price dynamics are shaped by both persistent causal drivers and time-localized responses, enhancing the interpretability and credibility of the proposed forecasting framework.

\subsubsection{Causality Examination in Autoformer}

Section~\ref{sec:causal_DAG} establishes causal DAGs that encode economically plausible structural relationships among decomposed price components, contract attributes, and exogenous market signals. This learned causal structure provides a reference for distinguishing variables that exert structural influence on retail electricity prices from those that act as non-causal temporal descriptors. While the proposed CG-TCN explicitly embeds this causal graph into its forecasting architecture, Autoformer is a purely data-driven transformer model that neither imposes nor learns an explicit causal structure, yet achieves competitive predictive accuracy and serves as a strong baseline in our empirical evaluation. In this subsection, we conduct a causal robustness examination of Autoformer by evaluating whether its predictive performance is consistent with the causal relationships identified in Section~\ref{sec:causal_DAG}, rather than being driven by non-causal temporal regularities.

Figures~\ref{fig:causal_graph_noconstraint}, \ref{fig:causal_graph_constraint}, and \ref{fig:causal_graph_nonlinear} show that day-of-week indicators (Day~0--Day~6) do not appear as causal parents of any forecasting target in the learned DAG, indicating that these calendar variables function as descriptive time indices rather than structural drivers of price formation. If Autoformer’s predictive accuracy were primarily grounded in causal relationships, perturbations to such non-causal calendar features would be expected to have limited impact on forecasting performance. To test this implication, we introduce a targeted calendar intervention that disrupts the alignment between day-of-week labels and observed prices and compare Autoformer’s forecasting accuracy before and after this intervention.

Operationally, we implement a within-week calendar shuffle that preserves all economic information and long-run temporal structure while selectively breaking intra-week calendar alignment. Starting from the chronologically ordered dataset, observations are partitioned into calendar weeks, and the date labels within each week are randomly permuted using a fixed random seed for reproducibility. This procedure preserves the set of observations within each week and leaves all price realizations, market variables, and contract attributes unchanged, while disrupting the mapping between specific weekdays (e.g., Monday through Sunday) and price outcomes. After permuting dates within each week, the dataset is re-sorted chronologically and used for evaluation. Because the intervention operates strictly within weekly blocks, it preserves seasonal patterns, long-term trends, and cross-week dependencies, thereby isolating the predictive contribution of within-week calendar regularities without introducing information leakage or altering the underlying price distribution.

\begin{table}[htbp]
\centering
\caption{Autoformer forecasting performance under within-week calendar intervention.}
\label{tab:autoformer_week_shuffle}
\small
\begin{tabular}{l r r r}
\toprule
Metric & Original data & \makecell{Within-week\\calendar--intervened data} & $\Delta$ (\%) \\ \midrule
MAE   & 0.266 & 0.346 & +30.1 \\
MSE   & 0.177 & 0.299 & +68.9 \\
RMSE  & 0.421 & 0.547 & +29.9 \\
MAPE  & 2.813 & 3.698 & +31.5 \\
NRMSE & 6.322 & 8.216 & +29.9 \\
$R^2$ & 0.961 & 0.934 & $-2.8$ \\
\bottomrule
\end{tabular}
\end{table}

Table~\ref{tab:autoformer_week_shuffle} summarizes Autoformer’s out-of-sample forecasting performance on the original dataset and under the within-week calendar intervention, together with the relative percentage change in each evaluation metric. The results show a substantial degradation in predictive accuracy following the intervention. Relative to the original evaluation, MAE increases by 30.1\%, RMSE by 29.9\%, and MAPE by 31.5\%, while MSE increases by 68.9\%. In parallel, $R^2$ declines from 0.961 to 0.934. Because the intervention disrupts only non-causal within-week calendar alignment while preserving all economic covariates, seasonal patterns, and long-run temporal structure, these performance losses indicate that a nontrivial portion of Autoformer’s predictive accuracy is driven by calendar-based proxies rather than structurally causal relationships. As a result, the feature--target associations learned by Autoformer are not causally interpretable in the sense of Section~\ref{sec:causal_DAG} and are not invariant to interventions that alter non-causal temporal regularities. This lack of invariance suggests potential fragility under policy interventions or regime changes that modify calendar effects, in contrast to CG-TCN, which explicitly constrains information flow through the learned causal structure and is therefore less reliant on such non-causal temporal descriptors.

\subsection{Ablation study}

To further validate the effectiveness of our proposed CG-TCN framework, we conducted an ablation study to investigate the contribution of each individual module. Specifically, we created variations of the full model by removing one component at a time, leading to the following configurations:

\begin{itemize}
    \item \textbf{w/o causal graph module}: The CG-TCN model without the causal graph discovery and graph neural network components. In this setup, the model only applies temporal convolution on the decomposed components, without incorporating causal relationships across variables. 
    \item \textbf{w/o TCN module}: The CG-TCN model without the temporal convolutional network. Instead of extracting temporal patterns via dilated convolution, the model uses a graph neural network for causal embedding followed by a multi-layer perceptron (MLP), which only utilizes the most recent time step. 
    \item \textbf{Pure TCN (univariate forecasting)}: A simplified baseline model in which the temporal convolutional network is trained only on the raw retail electricity price series, without any causal graph inputs or decomposition features. In this setting, the TCN captures purely univariate temporal dependencies of the price signal. It differs from the TCN baseline used before in our study: the baseline TCN forecasts each of the four decomposed price components and then aggregates them, whereas the pure TCN here directly forecasts the overall price trajectory without decomposition.
    \item \textbf{w/o spike-aware algorithm}: The CG-TCN model trained without the spike-aware mechanism, i.e., spike windows are not oversampled and no additional loss weighting is applied to spike-related errors.
\end{itemize}

Table~\ref{tab:ablation_results} reports the training and test performance of all ablation variants. Overall, the results highlight that each module contributes distinct and complementary benefits, and removing any single component leads to a measurable degradation in forecasting accuracy or generalization.

Removing the causal graph module leads to a clear decline in test performance, with MAE increasing to $0.5594$ and $R^2$ decreasing to $0.8391$. This indicates that explicitly modeling cross-variable causal relationships provides nontrivial predictive gains beyond temporal patterns alone, while also enhancing interpretability.

The absence of the temporal convolution module results in a more pronounced deterioration. Without the TCN’s ability to capture long-range temporal dependencies, test MAE increases to $0.9489$ and $R^2$ drops sharply to $0.6011$. This confirms that temporal convolution is essential for learning the dynamic structure of retail electricity prices.

The Pure TCN model performs worst among all variants, yielding a test MAE of $1.4339$ and a negative $R^2$ of $-0.1885$. This severe underperformance underscores the limitations of univariate forecasting in volatile retail electricity markets and highlights the importance of both multivariate causal information and decomposition-based feature representation.

Disabling the spike-aware algorithm still yields relatively strong average performance (test MAE = $0.2901$, $R^2 = 0.9463$), but performance is consistently inferior to the full CG-TCN model, particularly in terms of error metrics associated with high-volatility periods. This suggests that while the spike-aware mechanism is not the primary driver of baseline accuracy, it enhances robustness by improving sensitivity to rare but extreme price fluctuations.

\begin{table*}[htbp]
\centering
\caption{Ablation study results.}
\label{tab:ablation_results}
\resizebox{\textwidth}{!}{%
\begin{tabular}{
    l              % Model
    *6{c}          % 6 Training metrics
    *6{c}          % 6 Test metrics
}
\toprule
\textbf{Model} &
\multicolumn{6}{c}{\textbf{Training Results}} &
\multicolumn{6}{c}{\textbf{Test Results}} \\
\cmidrule(lr){2-7} \cmidrule(lr){8-13}
& \textbf{MAE} & \textbf{MSE} & \textbf{RMSE} &
  \textbf{MAPE} & \textbf{NRMSE} & $\mathbf{R^2}$ &
  \textbf{MAE} & \textbf{MSE} & \textbf{RMSE} &
  \textbf{MAPE} & \textbf{NRMSE} & $\mathbf{R^2}$ \\
\midrule
w/o causal graph module
 & 0.1623 & 0.0572 & 0.2393 &  2.42\% &  2.99\% &  0.9573
 & 0.5594 & 0.6189 & 0.7867 &  5.67\% & 14.16\% &  0.8391 \\

w/o TCN module
 & 0.3586 & 0.2742 & 0.5236 &  5.27\% &  6.54\% &  0.7956
 & 0.9489 & 1.5342 & 1.2386 & 10.25\% & 22.30\% &  0.6011 \\

Pure TCN
 & 1.0367 & 1.6677 & 1.2914 & 16.44\% & 16.14\% & -0.2430
 & 1.4339 & 4.5706 & 2.1379 & 13.94\% & 38.49\% & -0.1885 \\

w/o spike-aware algorithm
 & 0.0967 & 0.0548 & 0.2340 & 1.23\% & 2.93\% & 0.9624
 & 0.2901 & 0.1448 & 0.3733 & 3.43\% & 7.36\% & 0.9463 \\
\bottomrule
\end{tabular}
}
\end{table*}

In summary, the ablation study demonstrates that the causal graph module, temporal convolutional network, and spike-aware algorithm each play a critical and complementary role. Their integration enables CG-TCN to achieve accurate, stable, and volatility-aware forecasts, which is particularly important for decision-making in deregulated retail electricity markets.

\subsection{Sensitivity analysis}

Understanding how architectural choices affect forecasting performance is critical for deploying CG-TCN in practice, where model complexity, data availability, and computational cost must be carefully balanced. This sensitivity analysis systematically examines the impact of two key design parameters in the TCN module, the length of the historical input window and network depth, on predictive accuracy and generalization. Importantly, to isolate the effect of architecture alone, all models reported in this section are trained using the same fixed set of hyperparameters, without any additional hyperparameter optimization.

Table~\ref{tab:base_model_sensitivity} presents the results of this sensitivity analysis, where the input sequence length (10, 25, or 40 past steps) and the depth of the TCN (2 or 5 layers) are varied. Three main observations emerge.

First, the length of the historical input sequence has a pronounced effect on predictive performance. Models trained with only 10 past steps (e.g., 10step-2layerTCN) exhibit substantially higher test errors (MAE = 0.5563, $R^2$ = 0.8447) than those using longer input horizons. This suggests that retail electricity prices are influenced not only by short-term fluctuations but also by longer-term temporal dependencies. Extending the input window to 25 or 40 steps allows the model to better capture these slower-moving dynamics, resulting in more accurate forecasts.

Second, increasing network depth generally improves performance when sufficient historical context is available. For example, the 40step-2layerTCN achieves the best overall test performance, with the lowest MAE (0.3007), RMSE (0.4183), and MAPE (3.22\%), along with the highest $R^2$ (0.9528). This indicates that deeper or wider receptive fields enable the TCN to learn richer multi-scale temporal representations, which are well suited to the combination of volatility and structural persistence observed in retail electricity prices.

Finally, performance gains from increasing model complexity are not monotonic. While deeper networks can enhance representational capacity, they also introduce additional parameters and a higher risk of overfitting. For instance, the 25step-2layerTCN achieves competitive accuracy ($R^2$ = 0.9485) with substantially lower architectural complexity than deeper alternatives. This highlights a favorable trade-off between accuracy and efficiency, suggesting that moderate input lengths and depths may already capture most of the relevant temporal information.

\begin{table*}[htbp]
\centering
\caption{Sensitivity analysis of CG-TCN base models under different input window lengths and TCN depths (no hyperparameter optimization).}
\label{tab:base_model_sensitivity}
\resizebox{\textwidth}{!}{%
\begin{tabular}{
    l
    *6{c}
    *6{c}
}
\toprule
\textbf{Model} &
\multicolumn{6}{c}{\textbf{Training Results}} &
\multicolumn{6}{c}{\textbf{Test Results}} \\
\cmidrule(lr){2-7} \cmidrule(lr){8-13}
& \textbf{MAE} & \textbf{MSE} & \textbf{RMSE} &
  \textbf{MAPE} & \textbf{NRMSE} & $\mathbf{R^2}$ &
  \textbf{MAE} & \textbf{MSE} & \textbf{RMSE} &
  \textbf{MAPE} & \textbf{NRMSE} & $\mathbf{R^2}$ \\
\midrule
10step-2layerTCN
& 0.1518 & 0.0455 & 0.2133 & 2.32\% & 2.67\% & 0.9618
& 0.5563 & 0.6347 & 0.7967 & 5.62\% & 14.34\% & 0.8447 \\

10step-5layerTCN
& 0.1465 & 0.0330 & 0.1817 & 2.31\% & 2.27\% & 0.9754
& 0.4158 & 0.3487 & 0.5905 & 4.25\% & 10.63\% & 0.9093 \\

25step-2layerTCN
& 0.1356 & 0.0414 & 0.2034 & 2.04\% & 2.54\% & 0.9692
& 0.3664 & 0.1981 & 0.4451 & 4.16\% & 8.01\% & 0.9485 \\

25step-5layerTCN
& 0.1480 & 0.0334 & 0.1828 & 2.34\% & 2.28\% & 0.9751
& 0.4189 & 0.3536 & 0.5946 & 4.28\% & 10.70\% & 0.9081 \\

40step-2layerTCN
& 0.1497 & 0.0437 & 0.2092 & 2.26\% & 2.61\% & 0.9700
& \textbf{0.3007} & \textbf{0.1750} & \textbf{0.4183} & \textbf{3.22\%} & \textbf{7.53\%} & \textbf{0.9528} \\

40step-5layerTCN
& \textbf{0.1281} & \textbf{0.0295} & \textbf{0.1717} & \textbf{1.97\%} & \textbf{2.15\%} & \textbf{0.9797}
& 0.4002 & 0.2437 & 0.4936 & 4.56\% & 8.89\% & 0.9342 \\

\bottomrule
\end{tabular}
}
\end{table*}

Overall, the sensitivity analysis demonstrates that both temporal window size and network depth are influential architectural parameters for TCN-based forecasting. Longer input sequences provide richer historical context, while appropriate depth enhances temporal feature extraction. These results motivate the architectural choices adopted in the full CG-TCN model and underscore the importance of careful, problem-specific calibration rather than relying solely on increased model complexity.

\section{Discussion}

This study provides empirical and methodological evidence that incorporating causal structure into deep-learning–based forecasting can substantially improve both predictive performance and interpretability in deregulated retail electricity markets. Several broader insights emerge from the results.

First, the strong performance gains from integrating causal graphs with temporal convolution highlight the limitations of purely correlation-driven forecasting approaches in retail pricing contexts. Retail electricity prices are shaped by a combination of wholesale cost pass-throughs, contract design features, and regulatory interventions, whose effects unfold across multiple temporal scales. By explicitly modeling these relationships, the CG-TCN framework is better able to distinguish structural price movements from short-term noise, resulting in improved generalization, particularly in retail markets defined by price variability or volatility.

Second, the decomposition-based architecture proves critical for isolating heterogeneous temporal dynamics. The sensitivity and ablation analyses show that different price components respond differently to market drivers and architectural choices. Longer input windows and deeper temporal representations are especially valuable for capturing slower-moving trends linked to forward prices and contract terms, while the spike-aware mechanism enhances responsiveness to rare but impactful price fluctuations. These findings underscore the importance of aligning model design with the multi-scale nature of retail electricity price formation.

% Third, the learned causal graphs offer interpretable insights into price formation mechanisms that are difficult to obtain from standard black-box models. The results suggest that wholesale forward prices and contract attributes do not merely influence retail prices contemporaneously but propagate through specific temporal components, shaping both baseline price levels and large changes. Such insights are particularly relevant for regulators and consumer advocates seeking to understand how supplier pricing strategies translate into consumer-facing outcomes.

Third, the learned causal graphs offer interpretable insights into retail price formation mechanisms that are difficult to obtain from standard black-box models. The results indicate that wholesale forward prices primarily shape long-horizon trend components, reflecting cost pass-through and expectations embedded in upstream markets, while contract attributes such as monthly fees and early termination fees exert a stronger influence on short-term price fluctuations and discrete price adjustments. This pattern suggests that suppliers do not rely solely on per-kWh price levels to recover costs or generate revenue, but instead use fee-based contract features to modulate consumer-facing prices over shorter horizons. In particular, monthly fees and termination penalties appear to function as flexible instruments that allow suppliers to respond to market conditions and manage risk exposure without uniformly raising headline energy prices. Such insights are especially relevant for regulators and consumer advocates, as they highlight how supplier pricing strategies—including fee structures that may be less salient to consumers—translate into observed retail price dynamics.

% From a policy perspective, the framework provides a tool for anticipating periods of elevated price risk and assessing how market design features, such as default service rates or contract fee structures, may amplify or dampen retail price volatility. While this study focuses on a single utility territory, the methodology is transferable to other deregulated markets, offering a scalable approach for monitoring competition, affordability, and consumer exposure to price spikes.

From a policy perspective, the framework provides a tool for anticipating periods of elevated price risk and assessing how market design features, such as default service rates or contract fee structures, may influence retail price magnitudes or volatility. While this study demonstrates this robust methodology by focusing on depth rather than breadth (that is, on a single utility territory rather than all territories in the state), the methodology is generalizable to other deregulated markets, offering a scalable approach for developing reliable forecasts. This has implications for consumer protection, market monitoring, competition enforcement, welfare economic analysis and affordability assessment, and other regulatory and public-sector functions. It also has considerable private-sector applications, as the methods can be utilized by the many retail supply firms operating in these markets across the 14 US retail restructured states, in other retail choice markets such as natural gas choice markets, and internationally in other retail choice jurisdictions, including the UK and Australia.

It is also important to note another key contribution of this paper. Most empirical studies of electricity markets rely upon data from wholesale markets. Wholesale market data are free, abundant, and considerably less important from a public policy standpoint. This study utilizes a database of every daily retail choice offer from every supplier in a utility service territory, for a decade, in one of the largest and most robust retail electricity markets in the US. The data are costly to obtain in terms of effort, but considerably richer in behavioral detail and product heterogeneity than wholesale market data, which are more mechanistic. In short, we use data that are harder to obtain, richer and more robust, and ultimately more important and more relevant to real households and businesses because they reflect the prices they ultimately have to pay.

There are also many possible extensions of this methodology, as the framework is general and extensible. It offers a blueprint for causal forecasting in other volatile and structurally complex energy markets. First, extending CG-TCN toward probabilistic forecasting would enable estimation of the full distribution of daily available retail prices, moving beyond point forecasts. Such distributional predictions could support risk-aware procurement, budget planning, and consumer protection (e.g., identification of high price or risky offers that may endanger consumer welfare or low information consumers, such as the elderly). It could also support the quantification of uncertainty and tail risk during extreme market conditions. %Second, incorporating dynamic or time-varying causal discovery would allow the framework to adapt to regime shifts arising from policy changes, market restructuring, or macroeconomic shocks. 
Second, applying the proposed approach across multiple utility territories and contract types would enable cross-market generalization and comparative analysis, further strengthening the role of causal machine learning in energy system analytics.

%Finally, several limitations warrant attention. The causal discovery component relies on observed variables and obviously cannot account for unobservable factors such as differences in supplier-specific risk preferences, hedging behaviors (i.e., more versus less aggressive call options) or marketing strategies. In addition, while the current framework treats causal relationships as time-invariant, retail electricity markets may exhibit regime shifts due to policy changes or regional or national macroeconomic shocks. Addressing these limitations motivates several avenues for future research.

\section{Conclusion}

Retail electricity markets are the final downstream stage of the complex energy supply chain. All other levels of energy production and operations, from resource recovery to upstream and midstream operations, to wholesale markets, all culminate in the final end-product offered to households and businesses at the retail level. Retail electricity markets are of critical public policy importance, as production, regulatory and operational changes at all other stages of the supply chain are viewed through the lens of the markets that deliver the final product to the final consumer—households and job creators. 

Moreover, retail energy markets feature prominently in today’s charged political and policy debates in the US and many other nations, where energy affordability ranks in the top three policy issues for voters.  While historically the cost of utilities has been secondary or ancillary among public problems of concern, it has now risen to the third most highly cited affordability issue among Americans, surpassing even health care costs, and second only to housing and food. \footnote{ Boak, J. (2025, November 17). New analysis shows more US consumers are falling behind on their utility bills. AP News. https://apnews.com/article/trump-inflation-utility-bills-16cf846b44369b5b9660238be7112a4d} Consequently, methodological advancements in the prediction or forecasting of these important markets carry societal contributions above and beyond the empirical ones.

Retail electricity prices play a central role in shaping consumer decisions, supplier behavior, and overall market efficiency in deregulated retail energy markets. Accurately forecasting these prices requires models that go beyond short-term pattern recognition to capture the structural mechanisms underlying price formation. This paper introduced a causal-forecasting framework that integrates multi-resolution signal decomposition, causal graph discovery, and hybrid temporal deep learning for robust retail electricity price prediction.

By disentangling price dynamics across semiannual, quarterly, monthly, and high-frequency components and explicitly modeling their causal links with wholesale forward prices, contract attributes, and calendar effects, the proposed CG-TCN framework achieves superior forecasting accuracy relative to transformer-based and statistical benchmarks. Importantly, these gains are accompanied by improved interpretability, allowing the model to reveal how different market drivers propagate through distinct temporal layers of retail prices.

%Beyond predictive performance, this study demonstrates the value of embedding causal reasoning within modern forecasting architectures. The learned causal structures provide economically meaningful insights into the interaction between wholesale markets and retail contract design, offering a transparent view of how price levels and volatility emerge over time. Such interpretability is critical for regulators, policymakers, and market participants concerned with price stability, affordability, and consumer protection.

Beyond predictive performance, this study demonstrates the value of embedding causal reasoning within modern forecasting architectures. The learned causal structures provide economically meaningful insights into the interaction between wholesale markets and retail contract design, offering a methodology to improve the transparency of price formation in these behaviorally complex markets that have replaced a single regulated retail supplier with open market operations involving dozens of unregulated retail marketers or suppliers. Like other open market operations such as consumer credit, these marketers have a variety of contract attributes that serve as levers that they can modify or change (e.g., monthly fees, early termination fees, variable rates) to maximize profit beyond simply modifying the price or rate.  Understanding how wholesale and retail market characteristics contribute to retail price formation informs key economic and regulatory questions.  Advancements in interpretability are critical for regulators, policymakers, and market participants concerned with price stability, affordability, and consumer protection.

% Methodologically, the proposed framework is general and extensible, offering a blueprint for causal forecasting in other volatile and structurally complex energy markets. Several promising directions for future research emerge. First, extending CG-TCN toward probabilistic forecasting would enable estimation of the full distribution of daily available retail prices, rather than point forecasts alone. Such distributional predictions could support risk-aware procurement, budget planning, and consumer protection by quantifying uncertainty and tail risk during extreme market conditions. Second, incorporating dynamic or time-varying causal discovery would allow the framework to adapt to regime shifts arising from policy changes, market restructuring, or macroeconomic shocks. Finally, applying the proposed approach across multiple utility territories and contract types would enable cross-market generalization and comparative analysis, further strengthening the role of causal machine learning in energy system analytics.

Overall, this research advances retail electricity market forecasting by bridging accuracy and interpretability, delivering models that are not only predictive but also transparent and policy-relevant—an essential step toward more efficient, resilient, and consumer-oriented retail electricity markets.

\section{Acknowledgements}
This work was supported in part by the Alfred P. Sloan Foundation under Grant No. G-2022-19460. Any opinions, findings, and conclusions expressed are those of the authors and do not necessarily reflect those of the sponsor.

\newpage
\bibliographystyle{elsarticle-num}
%\bibliography{references, packages}
\bibliography{references}

\appendix  % Only once

% Appendix A
\renewcommand{\thefigure}{A.\arabic{figure}}
\setcounter{figure}{0}  % Reset figure counter if needed

\renewcommand{\thetable}{A.\arabic{table}}
\setcounter{table}{0}  % Reset table counter for B
\newpage
\section{The Potential Factors}
\label{sec:appendix_variables}

The potential factors are summarized in Tables~\ref{tab:contract_feature}, \ref{tab:energy_market}, and \ref{tab:time_feature}.

\begin{table}[htbp]
\centering
\caption{The retail electricity contract group.}
\small % or \footnotesize, \scriptsize
\resizebox{\textwidth}{!}{%
\begin{tabular}{cll}
\hline
\textbf{Index} & \textbf{Variables} & \textbf{Implication} \\
\hline
1  & MonthlyFee      & Fixed monthly fee included in the offer, in dollars. \\
2  & EarlyTerminationFee        & Fee charged for terminating the contract before it ends, in dollars. \\
\hline
\end{tabular}%
}
\label{tab:contract_feature}
\end{table}

\begin{table}[htbp]
\centering
\caption{The energy market group.}
\small % or \footnotesize, \scriptsize
\resizebox{\textwidth}{!}{%
\begin{tabular}{cll}
\hline
\textbf{Index} & \textbf{Variables} & \textbf{Implication} \\
\hline
1  & CalculatedFowardPrice      & \begin{tabular}[c]{@{}l@{}}Average forward wholesale electricity price for the forward maturity \\(delivery period) of a contract, using PJM West Hub prices \\for contracts originating on the date of the supply offer, in cents/kWh.\end{tabular} \\
2  & SSO        & \begin{tabular}[c]{@{}l@{}}Value in dollars per kWh of the standard service offer (SSO) prices \\on the day of the offer.\end{tabular} \\
\hline
\end{tabular}%
}
\label{tab:energy_market}
\end{table}

\begin{table}[htbp]
\centering
\caption{The time feature group.}
\small % or \footnotesize, \scriptsize
\resizebox{\textwidth}{!}{%
\begin{tabular}{cll}
\hline
\textbf{Index} & \textbf{Variables} & \textbf{Implication} \\
\hline
1  & Day\_0-6      & \begin{tabular}[c]{@{}l@{}} Contract publish day, from the first day to the end of each week.\end{tabular} \\
2  & Month\_1-12        & \begin{tabular}[c]{@{}l@{}} Contract publish month, from January to December each year.\end{tabular} \\
3  & Is\_holiday        & \begin{tabular}[c]{@{}l@{}} Contract publish date is a US holiday or not.\end{tabular} \\
\hline
\end{tabular}%
}
\label{tab:time_feature}
\end{table}

\subsection{Construction of Forward Wholesale Electricity Prices}
\label{sec:appendix_forward}

To control for the underlying cost of supplying energy for each retail contract offer, we construct a forward-looking wholesale electricity price that reflects the cost a supplier would face when hedging the contract on the day of origination for its full delivery horizon. This forward wholesale price aligns the maturity of wholesale forward contracts with the duration of each retail offer.

Specifically, for each retail contract \( i \) originating on day \( t \), with a contract length of \( n \) months and a publication month \( m \) in calendar year \( y \), we compute the average forward wholesale price as:

\begin{equation}
P^{(i)}_{\text{fwd}} =
\frac{1}{n}
\sum_{l=0}^{n-1}
P_{\text{wholesale}_{(m+l-1)\bmod 12 + 1,\; y + \left\lfloor \frac{m+l-1}{12} \right\rfloor}},
\end{equation}
where \( P_{\text{wholesale}_{t_1,t_2}} \) denotes the monthly wholesale electricity forward price for delivery in month \( t_1 \) of year \( t_2 \). The modulo operator ensures that contract months cycle correctly through the 12-month calendar, while the floor operator adjusts the delivery year when the contract horizon spans multiple calendar years.

This construction averages the sequence of monthly wholesale forward prices corresponding to the full delivery period of each retail contract, approximating the expected wholesale procurement cost that a supplier would incur if the contract were hedged at origination.

Wholesale forward prices are obtained from S\&P Market Intelligence for PJM’s West Hub, the most liquid trading hub in PJM and the proximate wholesale market for the Duke Energy Ohio service territory. Prices are expressed in cents per kilowatt-hour to maintain consistency with retail price outcomes and other market covariates.

% Appendix B
\renewcommand{\thefigure}{B.\arabic{figure}}
\setcounter{figure}{0}  % Reset figure counter if needed

\renewcommand{\thetable}{B.\arabic{table}}
\setcounter{table}{0}  
\newpage
\section{Time-Series Cross-Validation Strategy}
\label{sec:appendix_tscv}

In time-dependent forecasting tasks, traditional $k$-fold cross-validation—which randomly partitions the dataset—can lead to information leakage by allowing future observations to influence the training process. This violates the causal structure of time-series data, where past information alone should determine future outcomes. To maintain temporal causality, all experiments in this study employ a time-series cross-validation (TSCV) framework.

\begin{figure}[htbp]
  \centering
  \includegraphics[width=0.65\linewidth]{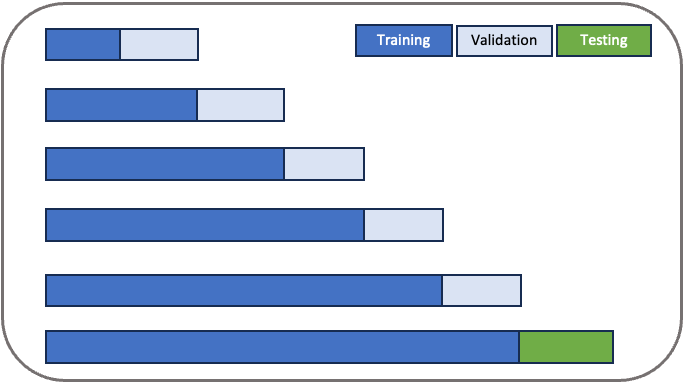}
      \caption{Illustration of the time-series cross-validation (TSCV) process. Each fold trains on historical observations (dark blue) and validates on the next temporal segment (light blue), rolling forward sequentially through time.}
  \label{fig:tscv_scheme}
\end{figure}

In TSCV, data are split sequentially so that each fold preserves chronological order. At the $i$-th fold, the model is trained on observations up to time $t_i$ and validated on the subsequent interval $(t_i, t_{i+1}]$. This ensures that the model only learns from historical data when forecasting unseen future values. The process is repeated by rolling or expanding the training window forward through time, as illustrated conceptually in Figure~\ref{fig:tscv_scheme}.

This design offers two key advantages:
\begin{enumerate}
    \item \textbf{Causality preservation.} By enforcing strict temporal order, the validation process respects real-world forecasting constraints, preventing the model from implicitly accessing future information.
    \item \textbf{Robustness assessment.} By testing performance across multiple historical windows, TSCV evaluates how well the model generalizes under changing market regimes and volatility conditions—an essential property for electricity price forecasting.
\end{enumerate}

Overall, the TSCV approach provides a more realistic and causally consistent evaluation of forecasting models compared to traditional random-split cross-validation, particularly in domains characterized by nonstationarity and regime shifts such as retail electricity pricing.

% Appendix C
\renewcommand{\thefigure}{C.\arabic{figure}}
\setcounter{figure}{0}  % Reset figure counter if needed

\renewcommand{\thetable}{C.\arabic{table}}
\setcounter{table}{0} 
\newpage
\section{Spike-Aware Algorithm}
\label{sec:appendix_spike}

This appendix details the spike-aware training mechanism used in the CG-TCN framework. The spike-aware algorithm is applied to the \emph{final reconstructed retail price series}, rather than to individual decomposed components, ensuring that extreme price movements are identified and treated consistently at the consumer-facing price level.

Spikes are defined as large positive or negative deviations from a smoothed baseline that exceed a specified $z$-score threshold. These spike windows capture short-lived but economically significant episodes of market stress and heightened volatility. Once identified, spike windows are used to guide both the data sampling strategy and the loss function during model training.

Because spike events occur infrequently, a naive training procedure may underrepresent them, leading to poor performance during extreme price movements. To address this imbalance, the spike-aware sampler oversamples spike windows with replacement such that, in expectation, a fixed proportion $p_{\text{spike}}$ of each mini-batch consists of spike observations. For example, when $p_{\text{spike}} = 0.5$, approximately half of each training batch is expected to contain spike windows, even though spikes account for a much smaller fraction of the full dataset. In parallel, a higher loss weight is assigned to prediction errors within spike windows, amplifying their influence on the training objective and improving the model’s responsiveness to abrupt price surges.

\begin{table}[htbp] 
\centering 
\caption{Number of spike windows identified in the final price series.} \label{tab:spike_windows} 
\begin{tabular}{lccc} 
\toprule \textbf{Fold} & 
\textbf{Training Windows} & \textbf{Spike Windows (Train)} & \textbf{Spike Windows (Valid)} 
\\ \midrule 1 & 598 & 123 / 598 & 92 / 597 
\\ 2 & 1,195 & 215 / 1,195 & 106 / 597 
\\ 3 & 1,792 & 321 / 1,792 & 120 / 597 
\\ 4 & 2,389 & 441 / 2,389 & 117 / 597 
\\ 5 & 2,986 & 558 / 2,986 & 109 / 597 
\\ \midrule \textbf{Total} & 3,583 & \textbf{667 / 3,583 (18.62\%)} & --- 
\\ 
\bottomrule 
\end{tabular} 
\end{table}

Table~\ref{tab:spike_windows} summarizes the number and proportion of spike windows identified across the full dataset and individual time-series cross-validation folds. This diagnostic serves as an important verification step, confirming that spike detection is neither overly aggressive nor overly conservative. By explicitly quantifying spike frequency, the spike-aware algorithm achieves a controlled balance between normal market conditions and extreme volatility episodes, improving robustness without distorting the underlying data distribution.

In total, 667 out of 3,583 windows (18.62\%) are identified as spike windows. These flagged periods are leveraged by the spike-aware sampler to balance training batches and improve learning under extreme price volatility, without artificially altering the underlying data distribution.

% Appendix D
\renewcommand{\thefigure}{D.\arabic{figure}}
\setcounter{figure}{0}  % Reset figure counter if needed

\renewcommand{\thetable}{D.\arabic{table}}
\setcounter{table}{0}  
\newpage
\section{Sensitivity of One-Sided Decomposition to Window Lengths}
\label{sec:appendix_window}

This appendix evaluates how the choice of decomposition window lengths affects downstream forecasting performance under one-sided STL and TSN decompositions. Because decomposition horizons directly control the smoothness, phase lag, and responsiveness of extracted components, their selection plays a critical role in balancing noise reduction against the preservation of predictive signals.

We consider multiple combinations of long-, medium-, and short-horizon windows,
\[
(H_L, H_M, H_S) \in \{(180,90,30), (180,60,20), (180,60,10), (180,90,15), (240,60,20)\},
\]
and apply each configuration consistently to both STL- and TSN-based one-sided decompositions. All results reported below correspond to one-step-ahead out-of-sample forecasts using identical model architectures and training procedures.

Table~\ref{tab:decomp_sensitivity_median} summarizes the forecasting performance for daily \emph{median} retail electricity prices. Among STL configurations, the $(180,60,20)$ setting consistently achieves the best accuracy, yielding the lowest MAE (0.2685), lowest RMSE (0.3609), and highest $R^2$ (0.9648). For TSN, shorter short-horizon windows are more effective, with the $(180,60,10)$ configuration providing the strongest overall performance ($R^2=0.9290$) and competitive error metrics. These results suggest that moderate medium-horizon smoothing combined with relatively short short-horizon components offers the best balance between stability and responsiveness for median price dynamics.

Table~\ref{tab:decomp_sensitivity_min} reports analogous results for daily \emph{minimum} prices, which are more volatile and spike-prone. In this setting, the optimal window choices differ slightly. For STL, the $(180,90,15)$ configuration performs best, indicating that a longer medium-term horizon helps stabilize extreme lower-bound price movements. For TSN, the $(180,60,10)$ configuration again delivers the strongest performance, achieving the lowest MAE (0.2772) and the highest $R^2$ (0.9693). Overall, the minimum-price results reinforce the importance of shorter short-horizon windows for capturing abrupt price movements, while highlighting that optimal medium-horizon smoothing may differ across price aggregates.

Based on these findings, the best-performing decomposition settings for median and minimum prices are adopted in the forecasting experiments.

\begin{table}[htbp]
\centering
\caption{One-step-ahead forecasting performance for retail electricity daily MEDIAN prices under different STL and TSN decomposition window settings.}
\label{tab:decomp_sensitivity_median}
\small
\begin{tabular}{l
                S[table-format=1.4]
                S[table-format=1.4]
                S[table-format=1.4]
                S[table-format=3.2, table-space-text-post=\%]
                S[table-format=2.2, table-space-text-post=\%]
                S[table-format=1.4]}
\toprule
\multicolumn{1}{c}{\textbf{Method / Window}} &
\multicolumn{1}{c}{\textbf{MAE}} &
\multicolumn{1}{c}{\textbf{MSE}} &
\multicolumn{1}{c}{\textbf{RMSE}} &
\multicolumn{1}{c}{\textbf{MAPE}} &
\multicolumn{1}{c}{\textbf{NRMSE}} &
\multicolumn{1}{c}{$\mathbf{R^2}$} \\
\midrule
STL\_180-90-30    & 0.3685 & 0.3051 & 0.5524 & 3.98\%  & 9.94\%  & 0.9176 \\
\textbf{STL\_180-60-20} & \textbf{0.2685} & \textbf{0.1302} & \textbf{0.3609} & \textbf{3.08\%} & \textbf{6.49\%} & \textbf{0.9648} \\
STL\_180-60-10     & 0.6992 & 0.7170 & 0.8468 & 7.82\% & 15.24\% & 0.8063 \\
STL\_180-90-15     & 0.8502 & 1.2869 & 1.1344 & 9.90\% & 20.42\% & 0.6525 \\
STL\_240-60-20     & 0.8604 & 1.2138 & 1.1017 & 9.95\% & 19.83\% & 0.6722 \\
\midrule
TSN\_180-90-30     & 0.4850 & 0.3543 & 0.5952 & 5.84\% & 10.71\% & 0.9043 \\
TSN\_180-60-20     & 0.5728 & 0.7569 & 6.3677 & 6.36\% & 13.62\% & 0.8453 \\
\textbf{TSN\_180-60-10} & \textbf{0.4202} & \textbf{0.2630} & \textbf{0.5129} & \textbf{5.01\%} & \textbf{9.23\%} & \textbf{0.9290} \\
TSN\_180-90-15     & 0.4006 & 0.2639 & 0.5137 & 4.54\% & 9.24\% & 0.9287 \\
TSN\_240-60-20     & 0.4899 & 0.3579 & 0.5982 & 5.68\% & 10.76\% & 0.9034 \\
\bottomrule
\end{tabular}
\end{table}

\begin{table}[htbp]
\centering
\caption{One-step-ahead forecasting performance for retail electricity daily MINIMUM prices under different STL and TSN decomposition window settings.}
\label{tab:decomp_sensitivity_min}
\small
\begin{tabular}{l
                S[table-format=1.4]
                S[table-format=1.4]
                S[table-format=1.4]
                S[table-format=2.2, table-space-text-post=\%]
                S[table-format=1.4]}
\toprule
\multicolumn{1}{c}{\textbf{Method / Window}} &
\multicolumn{1}{c}{\textbf{MAE}} &
\multicolumn{1}{c}{\textbf{MSE}} &
\multicolumn{1}{c}{\textbf{RMSE}} &
\multicolumn{1}{c}{\textbf{NRMSE}} &
\multicolumn{1}{c}{$\mathbf{R^2}$} \\
\midrule
STL\_180-90-30 & 1.0907 & 2.2566 & 1.5022 & 15.50\% & 0.5488 \\
STL\_180-60-20 & 0.8240 & 1.3744 & 1.1723  & 12.09\% & 0.7252 \\
STL\_180-60-10 & 0.8700 & 1.9127 & 1.3830  & 14.27\% & 0.6176 \\
\textbf{STL\_180-90-15} 
& \textbf{0.4871} 
& \textbf{0.4497} 
& \textbf{0.6706} 
& \textbf{6.92\%} 
& \textbf{0.9101} \\
STL\_240-60-20 & 1.1898 & 2.2200 & 1.4900 & 12.88\% & 0.5709 \\
\midrule
TSN\_180-90-30 & 0.4198 & 0.3961 & 0.6294 & 6.49\% & 0.9208 \\
TSN\_180-60-20 & 0.5492 & 0.4892 & 0.6994 & 7.21\% & 0.9022 \\
\textbf{TSN\_180-60-10} 
& \textbf{0.2772} 
& \textbf{0.1534} 
& \textbf{0.3917} 
& \textbf{4.04\%} 
& \textbf{0.9693} \\
TSN\_180-90-15 & 0.3567 & 0.3085 & 0.5554  & 5.73\% & 0.9383 \\
TSN\_240-60-20 & 0.4072 & 0.5831 & 0.7636  & 7.88\% & 0.8834 \\
\bottomrule
\end{tabular}
\end{table}

% Appendix E
\renewcommand{\thefigure}{E.\arabic{figure}}
\setcounter{figure}{0}  % Reset figure counter if needed

\renewcommand{\thetable}{E.\arabic{table}}
\setcounter{table}{0}  
\newpage
\section{Hyperparameter Optimization}
\label{sec:appendix_parameter}

This appendix describes the hyperparameter optimization procedure used for the proposed CG-TCN model and two strong deep-learning baselines—MTGNN and Autoformer. The goal of this procedure is to ensure a fair and systematic comparison by tuning each model within its standard architectural design while avoiding ad hoc manual calibration.

\subsection{Optuna-based optimization framework}

We employ Optuna \cite{akiba2019optuna}, an open-source framework for automated hyperparameter optimization, to search the parameter spaces of CG-TCN, MTGNN, and Autoformer. Optuna adopts a \emph{define-by-run} paradigm, allowing flexible specification of model-specific search spaces that include both architectural parameters (e.g., network depth and channel dimensions) and training-related parameters (e.g., learning rate, dropout, and weight decay).

The optimization process is guided by the Tree-structured Parzen Estimator (TPE) algorithm \cite{bergstra2013making}, which adaptively balances exploration and exploitation when sampling candidate configurations. To improve computational efficiency, Optuna’s pruning mechanism is enabled to terminate unpromising trials early based on intermediate validation performance. This strategy substantially reduces training cost while mitigating overfitting to specific hyperparameter choices.

For each model, the objective function minimizes the validation loss computed on a held-out subset of the training data using time-series cross-validation. The best-performing configuration identified by Optuna is then retrained on the full training set and evaluated on the test set. The resulting optimal hyperparameters are reported in Tables~\ref{tab:causal_tcn_best_params}--\ref{tab:autoformer_best_params}.

For the proposed CG-TCN model, the hyperparameter search spans learning rate, batch size, dropout rates for the TCN, GNN, and fusion layers, kernel size, number of TCN layers, channel widths, and GNN output dimensions. In addition, parameters specific to the spike-aware mechanism—such as the spike detection threshold, spike loss weight, and spike window sampling probability—are jointly optimized to ensure robust performance during extreme price events. The complete set of selected hyperparameters is reported in Table~\ref{tab:causal_tcn_best_params}.

\begin{table}[htbp]
\centering
\caption{Optimal hyperparameters for CG-TCN.}
\label{tab:causal_tcn_best_params}
\small
\begin{tabular}{l S[table-format=1.6]}
\toprule
\textbf{Parameter} & \multicolumn{1}{c}{\textbf{Value}} \\
\midrule
\texttt{learning\_rate}          & 0.0002315379549016443 \\
\texttt{batch\_size}             & 16 \\
\texttt{dropout\_tcn}            & 0.35164762264507865 \\
\texttt{dropout\_gnn}            & 0.20361823737646584 \\
\texttt{dropout\_fusion}         & 0.12467108935051119 \\
\texttt{kernel\_size\_tcn}       & 3 \\
\texttt{num\_tcn\_layers}        & 2 \\
\texttt{tcn\_channels\_l0}       & 16 \\
\texttt{tcn\_channels\_l1}       & 32 \\
\texttt{gnn\_output\_dim}        & 8 \\
\texttt{weight\_decay}           & 1.3768096295736596e-06 \\
\texttt{noise\_std}              & 0.040450432581941684 \\
\texttt{spike\_z}                & 2.3143109433972024 \\
\texttt{loss\_w\_spike}          & 5.420188496403048 \\
\texttt{sampler\_p\_spike}       & 0.48620254590795026 \\
\texttt{num\_epochs\_hpo}        & 122 \\
\texttt{early\_stopping\_patience} & 20 \\
\bottomrule
\end{tabular}
\end{table}

For MTGNN, the search space includes architectural parameters governing temporal convolutions and graph learning, such as the number of layers, graph convolution depth, subgraph size, channel dimensions, and dilation factors. Training-related parameters, including learning rate, dropout, batch size, and gradient clipping, are also optimized. The optimal configuration identified by Optuna is summarized in Table~\ref{tab:mtgnn_best_params}.

\begin{table}[htbp]
\centering
\caption{Optimal hyperparameters for MTGNN.}
\label{tab:mtgnn_best_params}
\small
\begin{tabular}{l S[table-format=1.6]}
\toprule
\textbf{Parameter} & \multicolumn{1}{c}{\textbf{Value}} \\
\midrule
\texttt{learning\_rate}        & 0.0001025350969016849 \\
\texttt{weight\_decay}         & 0.006351221010640704 \\
\texttt{dropout}               & 0.43919636508684307 \\
\texttt{batch\_size}           & 16 \\
\texttt{seq\_in\_len}           & 96 \\
\texttt{layers}                & 8 \\
\texttt{gcn\_depth}            & 3 \\
\texttt{subgraph\_size}        & 12 \\
\texttt{node\_dim}             & 64 \\
\texttt{dilation\_exponential} & 2 \\
\texttt{conv\_channels}        & 8 \\
\texttt{residual\_channels}    & 16 \\
\texttt{skip\_channels}        & 64 \\
\texttt{end\_channels}         & 96 \\
\texttt{propalpha}             & 0.06787661614294042 \\
\texttt{tanhalpha}             & 1.4883605700319193 \\
\texttt{num\_split}            & 1 \\
\texttt{step\_size}            & 200 \\
\texttt{clip}                  & 5 \\
\bottomrule
\end{tabular}
\end{table}

For Autoformer, hyperparameter optimization focuses on transformer-specific architectural choices, including embedding dimension, number of attention heads, encoder and decoder depths, feed-forward network width, and moving-average window length used in the series decomposition module. Standard training parameters such as learning rate, dropout, and batch size are optimized jointly. The selected hyperparameters are reported in Table~\ref{tab:autoformer_best_params}.

\begin{table}[htbp]
\centering
\caption{Optimal hyperparameters for Autoformer.}
\label{tab:autoformer_best_params}
\small
\begin{tabular}{l S[table-format=3.6]}
\toprule
\textbf{Parameter} & \multicolumn{1}{c}{\textbf{Value}} \\
\midrule
\texttt{d\_model}        & 256 \\
\texttt{n\_heads}        & 4 \\
\texttt{e\_layers}       & 3 \\
\texttt{d\_layers}       & 1 \\
\texttt{d\_ff}           & 512 \\
\texttt{batch\_size}     & 16 \\
\texttt{learning\_rate}  & 0.0005498966977327428 \\
\texttt{dropout}         & 0.17806470410412306 \\
\texttt{moving\_avg}     & 25 \\
\texttt{factor}          & 3 \\
\texttt{activation}      & {gelu} \\
\bottomrule
\end{tabular}
\end{table}

\subsection{The other baseline models}

For the remaining baseline models—LSTM, TCN, and CTCN—we do not perform automated hyperparameter optimization. Instead, these models are trained using commonly adopted configurations from the literature to serve as reference benchmarks rather than fully optimized competitors. All baseline models are trained using the Adam optimizer \cite{kingma2015adam}, which combines adaptive learning rates with momentum-based updates.

Adam maintains exponential moving averages of the first- and second-order moments of the gradients to scale parameter updates adaptively, leading to faster convergence and improved stability compared to vanilla stochastic gradient descent. In our implementation, Adam is configured with weight decay for learning rate regularization, while all other optimizer settings follow the default values in PyTorch.

% Appendix F
\renewcommand{\thefigure}{F.\arabic{figure}}
\setcounter{figure}{0}

\renewcommand{\thetable}{F.\arabic{table}}
\setcounter{table}{0}

\newpage
\section{Comparison with the Causality from CTCN Model}
\label{sec:appendix_ctcn_comparison}

This appendix contrasts the causal interpretation obtained from the baseline CTCN model with that of the proposed CG--TCN framework, highlighting fundamental differences in how predictive influence and causality are represented.

\begin{figure}[htbp]
  \centering
  \includegraphics[width=0.9\linewidth]{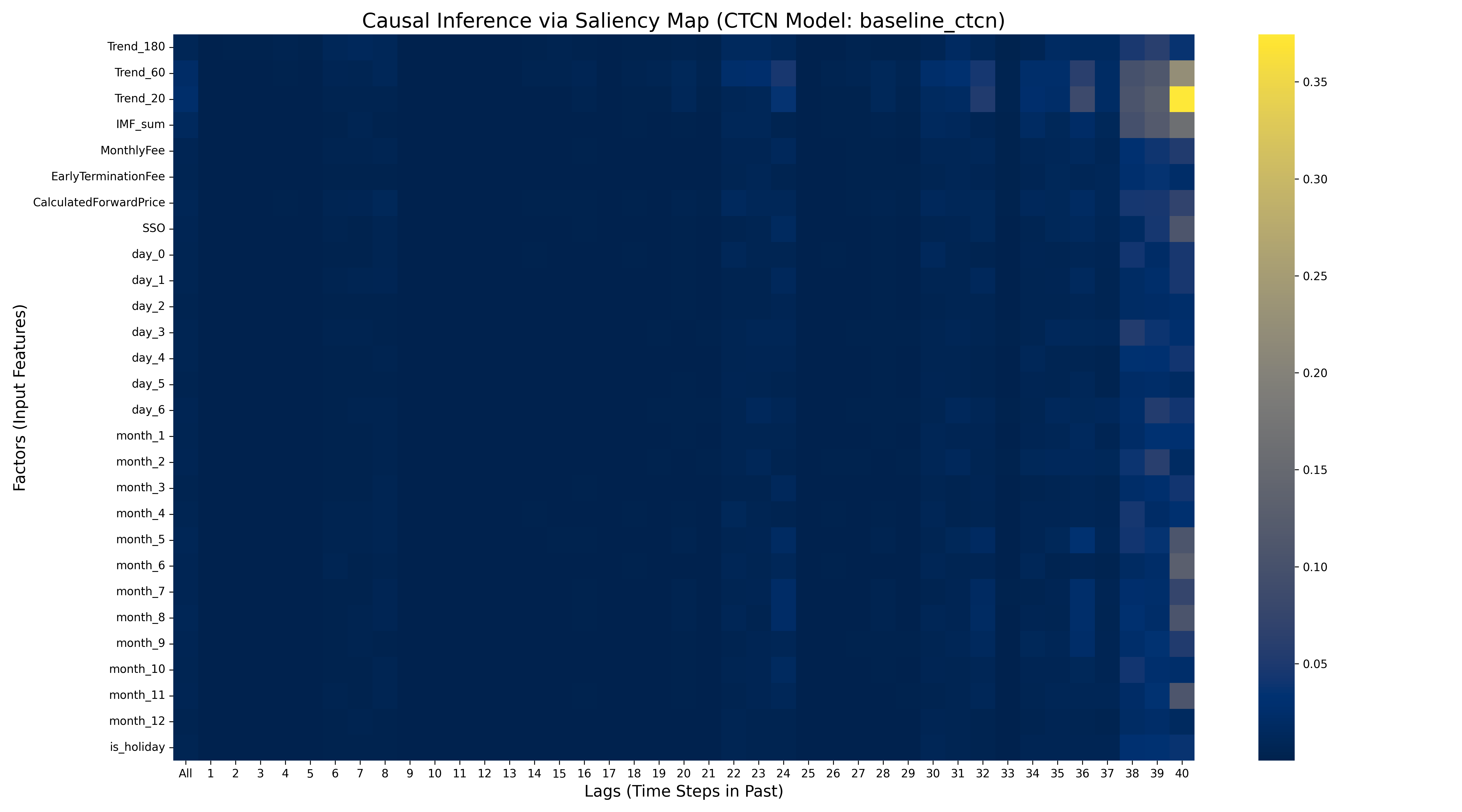}
  \caption{Saliency-based causal inference results from the baseline CTCN model. The model places dominant emphasis on long-term trend components while underutilizing exogenous or calendar-based features.}
  \label{fig:causal_ctcn}
\end{figure}

Figure~\ref{fig:causal_ctcn} presents saliency-based attribution results for the baseline CTCN. The heatmap reports the relative importance of input features across historical lags, reflecting how the model assigns predictive influence through time. The CTCN places dominant emphasis on the long-horizon trend component (Trend\_180), particularly at distant lags, suggesting that slowly evolving structural patterns account for most of the model’s predictive power. In contrast, shorter-horizon components (Trend\_60, Trend\_20, and IMF\_sum), contract attributes (MonthlyFee, EarlyTerminationFee), and calendar indicators (day-of-week, month, and holiday effects) receive comparatively little weight. While this behavior indicates that the CTCN effectively captures temporal persistence, the saliency patterns alone cannot distinguish true causal drivers from features that are merely correlated with long-term trends.

A comparison with the proposed CG--TCN framework reveals a key conceptual distinction. The CTCN adopts a \emph{flat attribution structure}, in which all covariates directly contribute to the retail price prediction. Although this approach can capture strong statistical associations, it risks conflating correlation with causation, particularly in settings where multiple inputs share common temporal structure. By contrast, CG--TCN embeds an \emph{explicit causal graph} that organizes dependencies hierarchically, routing the influence of exogenous variables through decomposed price trends before affecting the final retail price. This design redistributes attribution across interpretable causal pathways rather than concentrating weight on a small set of dominant predictors.

\begin{figure}[htbp]
  \centering
  \begin{subfigure}[t]{0.48\textwidth}
    \centering
    \includegraphics[width=\linewidth]{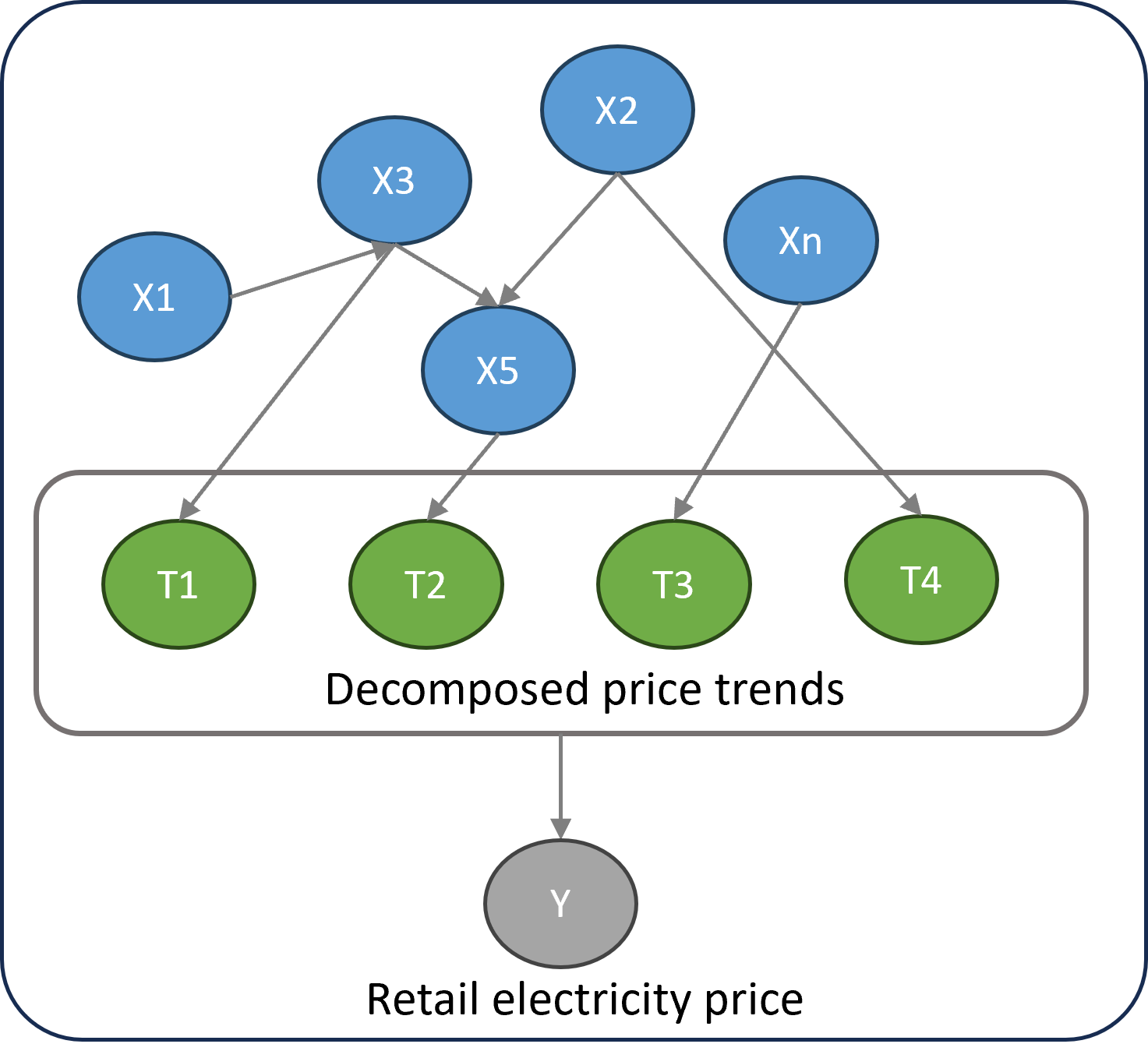}
    % \caption{CG–TCN: hierarchical causal pathways (exogenous → decomposed trends → price).}
    \label{fig:causality_cgtcn}
  \end{subfigure}
  \hfill
  \begin{subfigure}[t]{0.48\textwidth}
    \centering
    \includegraphics[width=\linewidth]{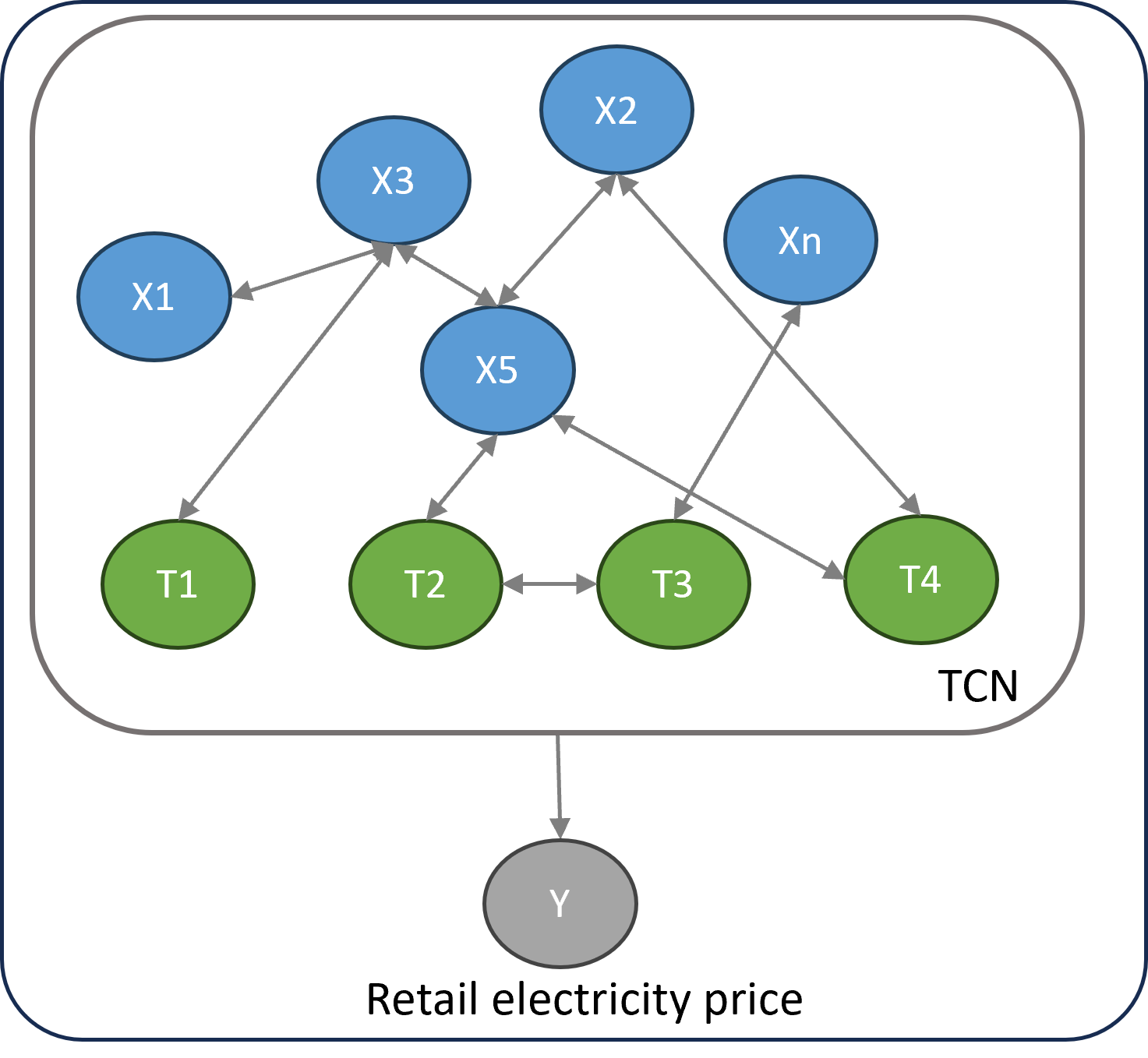}
    % \caption{CTCN: flat structure with all covariates directly predicting price.}
    \label{fig:causality_ctcn}
  \end{subfigure}
  \caption{Conceptual comparison of causal structures. 
  (a) The CG–TCN encodes a hierarchical causal graph that channels exogenous effects through decomposed price trends; 
  (b) The baseline CTCN treats all inputs as direct predictors, potentially conflating correlations with true causal drivers.}
  \label{fig:causality_comparison}
\end{figure}

Figure~\ref{fig:causality_comparison} illustrates this difference schematically. In the CTCN representation (panel~b), decomposed trends and exogenous variables are treated as independent predictors of retail electricity prices, without modeling their causal interdependencies. Such a structure may absorb exogenous effects into dominant trend components, obscuring the underlying price-formation mechanisms. In contrast, the CG--TCN (panel~a) enforces a layered causal hierarchy: exogenous features first influence intermediate price trends through learned causal links, which then jointly determine the final retail price. This hierarchical formulation yields a more faithful and interpretable representation of retail electricity price dynamics, grounding predictive performance in structured causal reasoning rather than purely empirical correlations.

\end{document}